\documentclass[12pt,twoside]{article}

\usepackage{amssymb}
\usepackage{amsmath}
\usepackage{latexsym}
\usepackage{longtable}
\usepackage{epsfig}
\usepackage{graphicx,bm,psfrag}

\newcommand{\rme}{\mathrm{e}}
\newcommand{\ri}{\mathrm{i}}
\renewcommand{\leq}{\leqslant}
\renewcommand{\geq}{\geqslant}

\begin{document}
\begin{titlepage}
\begin{center}
\vspace*{2cm} {\Large {\bf Nonlinear Fluctuating Hydrodynamics in One Dimension: the Case of Two Conserved Fields} \bigskip\bigskip\\}
{\large Herbert Spohn$^1$ and Gabriel Stoltz$^2$}
\bigskip
\bigskip\\
1: Zentrum Mathematik and Physik Department, TU M\"unchen,\\
Boltzmannstr. 3, D-85747 Garching, Germany\\e-mail: {\tt spohn@ma.tum.de}\\
\bigskip
2: Universit\'e Paris-Est, CERMICS (ENPC), INRIA \\
F-77455 Marne-la-Vall\'ee, France \\
email: {\tt stoltz@cermics.enpc.fr}        
\end{center}
\vspace{5cm} 
\textbf{Abstract.} We study the BS model, which is a one-dimensional lattice field theory taking real values. Its dynamics is governed by coupled differential equations plus random nearest neighbor exchanges. The BS model has exactly two locally conserved fields. Through numerical simulations the peak structure of the steady state space-time correlations is determined  and compared with nonlinear fluctuating hydrodynamics, which predicts a traveling peak with KPZ scaling function and a standing peak with a scaling function given by the completely asymmetric Levy distribution with parameter $\alpha = 5/3$. As a by-product, we  completely classify the universality classes for two coupled stochastic Burgers equations with arbitrary coupling coefficients.  

\end{titlepage}

%-------------------------------------------------------
\section{Introduction}
\label{sec1}
\setcounter{equation}{0}

As recognized for some time \cite{ErHa76,FoNe77,Livi97,LeLi03,Dh08,vB12}, one-dimensional systems generically
have anomalous transport properties. They can be observed through the super-diffusive spreading of  small perturbations in a homogeneous steady state. An alternative, equally popular route is to consider a system of finite length, $L$, and to impose a fixed difference in the value of conserved fields at the two boundary points.  For regular transport the resulting steady state current behaves as $L^{-1}$, while anomaly means an enhanced current of order $L^{-1 + \tilde{\alpha}}$ with some $\tilde{\alpha} > 0$.

Recently it has been proposed that such anomalous transport could be understood through a nonlinear extension of fluctuating hydrodynamics~\cite{Spohn14}. This method presupposes the availability of locally conserved fields, say $n$ in total number, where $n = 1,2,3$ mostly. The dynamics can be quite general, classical, quantum, stochastic, under the restriction of being translation invariant and having sufficiently local interactions. The basic construction is easily explained:
One first has to identify all locally conserved fields, $n$ of them. Integrable systems are thereby ruled out because their number of conserved fields is proportional to system size. The dynamics admits then an $n$-parameter family of translation invariant  steady  states. For them one has to compute the steady state average currents, which thus are functions
of the steady state average of the conserved fields. In order to have anomalous behavior these macroscopic current functions have to be nonlinear. There are models in which the currents are identically zero (or linear), which is then a strong indication for regular, diffusive transport.

Even if the current functions are nonlinear, there are still several distinct universality classes. To systematically explore their structure is  one goal of our contribution. While one could consider the general case of $n$ conserved fields, it seems to us more instructive to stick to the simplest case of $n=2$, which already exhibits the main mechanisms at work.
The one-component case has been studied in great detail under the heading of one-dimensional Kardar-Parisi-Zhang (KPZ) equation \cite{KPZ}, see also  the recent reviews \cite{Co12,BG12,BP13,QR13}.  We will make use of these results, but our focus is on the novel features arising for $n =2$. 

In the following we will consider only the spreading of small perturbations, which is identical to investigating the steady state space-time correlations of the conserved fields. (The issue of steady states with open boundaries remains as a challenge for the future.) We will work out their scaling behavior on the basis of nonlinear fluctuating hydrodynamics, partially exact,
partially approximate, and provide a complete classification of the universality classes for $n =2$.

Such results are of  interest only when compared with microscopic models which are accessible through numerical simulations. This still leaves a very wide choice, but there are constraints. Firstly it is convenient to have a lattice type
model. In addition, the steady states should be explicit. In the most favorable cases the steady states are of product form, for which static averages are then easily obtained. As an aside, thereby one can also compute explicitly the non-universal coefficients, making our predictions more pointed. Of course, we also would like  the microscopic model  to have features which have not been observed before. Our choice here is a one-dimensional lattice field theory taking real values. The dynamics is given by a system of coupled differential equations and admits two locally conserved quantities. To have good space-time mixing, and to avoid spurious conservation laws, we add a simple stochastic exchange term. Our model has been first introduced  
in~\cite{BS12}. Its novel feature is the two peak structure for the steady state correlation functions, one of them travels with  
a strictly negative velocity and has KPZ scaling, while the other one is standing still and has a scaling function given by the Levy distribution
with parameter $\alpha = 5/3$ and maximal asymmetry $b = 1$. Numerical simulations are performed for two different potentials, the exponential potential and an asymmetric FPU potential.

To provide a brief outline, in Section \ref{sec2} we discuss the universality classes for a generic two-component system
in the framework of nonlinear fluctuating hydrodynamics. The microscopic model is introduced in Section \ref{sec3},
while the results of the respective numerical solutions are reported in Section \ref{sec:numerics}. Extra material, requiring more lengthy computations, is shifted to the Appendices.\\\\
\textbf{Acknowledgements}.  We thank Christian Mendl for numerous instructive discussions, as well as G\"unther Sch\"utz for stimulating comments on a preliminary version of this manuscript. H.S. is grateful for the support through the Institute for Advanced Study, Princeton, where the first steps in this project were accomplished and thanks David Huse for insisting on a complete classification. 

%-------------------------------------------------------
\section{Two-component stochastic Burgers equation}
\label{sec2}
\setcounter{equation}{0}

Stochastic Burgers equations are a convenient way to formulate systems of hyperbolic conservation laws including noise.
The nonlinearity of the systematic currents are kept to quadratic order. A linear dissipative term is also included. All other degrees of freedom are subsumed as fluctuating currents, for simplicity  modeled as space-time white noise. 
The resulting system of stochastic conservation laws is somewhat singular \cite{Ha13}, but  extremely useful in classifying the various universality classes. Applications concern suitably discrete versions.
In our contribution we restrict ourselves to the case of two components, which already illustrates  well the main features of systems with an arbitrary number of components.

\subsection{One-component systems}
\label{sec:one_component}

Let us first briefly recall the case of a single component, $u_1(x,t)$, which by assumption is governed by the stochastic Burgers equation
\begin{equation}\label{2.1}
\partial_tu_1+\partial_x\big(c u_1 + G^1_{11}u^2_1 -D\partial_x u_1 +\sqrt{2D}\xi_1\big)=0\,,
\end{equation}
where $D> 0$ is the viscosity, $c \in \mathbb{R}$ the  velocity of propagation, $G_{11}^1 \in \mathbb{R}$ the strength of the nonlinearity, and $\xi_1$  a space-time white noise of unit strength. (We use redundant notation, as $u_1$, 
$G^1_{11}$ to be in accord with the  case of two components). We are interested in the stationary process governed by~(\ref{2.1}). As proved in \cite{FuQu14}, spatial white noise with mean zero and unit variance  is an invariant measure for (\ref{2.1}). The most basic object of real interest is then the stationary space-time covariance $\langle u_1(x,t)u_1(0,0)\rangle$, where
$\langle \cdot \rangle$ refers to the expectation with respect to the stationary process. Recently an exact solution has been accomplished \cite{BoCoFe14a}, which turns out to validate prior non-rigorous replica computations \cite{ImSa12}.  
For large $x,t$ the exact solution behaves  as
\begin{equation}\label{2.2}
\langle u_1(x,t)u_1(0,0)\rangle\simeq (\lambda_\mathrm{B} t)^{-2/3} f_{\mathrm{KPZ}}
\left((\lambda_\mathrm{B} t)^{-2/3}(x -ct)\right)\,,
\end{equation}
where $\lambda_{\mathrm{B}}= 2\sqrt{2}|G^1_{11}|$. The universal scaling function $f_{\mathrm{KPZ}}$ is tabulated in~\cite{Prhp}, there denoted by $f$, and has the following properties: $f_{\mathrm{KPZ}}\geq 0$, $f_{\mathrm{KPZ}}(x)=f_{\mathrm{KPZ}}(-x)$, 
\[
\int_\mathbb{R} f_{\mathrm{KPZ}}(x) \, dx = 1, \qquad \int_\mathbb{R} f_{\mathrm{KPZ}}(x) \, x^2 \, dx =0.510523\ldots\,.
\]
In fact, $f_{\mathrm{KPZ}}$ looks roughly like a Gaussian distribution but with faster decaying tails  as $\exp(-0.295|x|^{3})$, see~\cite{PrSp04}. The $f_{\mathrm{KPZ}}$ scaling behavior for the stationary two-point function has been proved also for the PNG model \cite{PrSp04}, the TASEP \cite {FeSp06}, and the semi-discrete directed polymer model \cite{BoCoFe14a} and is expected to be valid for the entire KPZ universality class.

\subsection{Classification of two-component systems}
\label{sec:complete_classification}

Let us turn to the case of two components $\vec{u} = (u_1,u_2)$. The coupling constants become matrices and it is of advantage to stick to the most general form which reads
\begin{equation}\label{2.3}
\partial_tu_\alpha+\partial_x\big(c_\alpha u_\alpha  + \vec{u} \cdot G^\alpha \vec{u}  -\partial_x (D\vec{u})_\alpha + (\sqrt{2D}\vec{\xi}\,)_\alpha\big)=0\,, \qquad \alpha = 1,2,
\end{equation}
where $c_\alpha$ is the propagation velocity of the $\alpha$-th component, the symmetric matrices $G^\alpha \in \mathbb{R}^{2 \times 2}$ determine the strength of the nonlinearity, the diffusion matrix $D \in \mathbb{R}^{2 \times 2}$ is symmetric positive, and $\vec{\xi}$ is a vector of two independent mean zero Gaussian white noises with covariance $\langle \xi_\alpha (x,t) \xi_{\alpha'}(x',t')\rangle = \delta_{\alpha\alpha'} \delta(x-x')\delta(t- t')$. Note that (\ref{2.3}) is written already in normal coordinates, which are defined by the  linear drift part of the current being diagonal,
see~\cite{Spohn14} for more precision as well as~\eqref{eq:normal_mode_transform} below. In (\ref{2.1}) the 
term $cu_1$ can be removed by switching to a coordinate system moving with velocity $c$. Under the same transformation, for a two-component system the relative velocity necessarily persists, which is the origin for much richer properties. 

As before, our interest is in the stationary process governed by~(\ref{2.3}), in particular its covariance matrix $\langle u_\alpha(x,t)u_{\alpha'}(0,0)\rangle$.  No exact solutions are available and we have to work with approximations. The first issue is already the invariant measure of~(\ref{2.3}). Only if $G^1_{22} = G^2_{12}$ and 
$G^2_{11} = G^1_{12}$, the invariant measure is known to be white noise in $x$ with independent components.
Our choice of the noise strength ensures unit strength for both components.
The linear case, $G^1=0 = G^2$, is easily solved.   If $c_1 \neq c_2$, then for large $x,t$, the covariance consists of  two decoupled Gaussian peaks, respectively centered at $c_\alpha t$ and of width $\sqrt{D_{\alpha\alpha} t}$. Note that possible cross terms of $D$ do not show up, since the peaks move with distinct velocities. To a certain extent, this feature will still be valid, once the nonlinearity is included. Hence we will assume 
$c_1 \neq c_2$ throughout. The case $c_1 = c_2$ has to be studied separately, see \cite{EK93} for an early discussion.

Since the mode velocities differ, the linear drift term is dominant and one expects that in general  the two equations in~\eqref{2.3} decouple for large $x,t$. However, if one of the leading non-linear couplings, $G^\alpha_{\alpha\alpha}$, vanishes, the argument becomes more subtle. To gain some insight we turn directly to the mode-coupling approximation for (\ref{2.3}).  It is based on a suitable Gaussian approximation together with the observation that the off-diagonal terms of the covariance are very small, see~\cite[Appendix~C]{Spohn14}. More precisely,
\begin{equation}\label{2.4}
\langle u_\alpha(x,t)u_{\alpha'}(0,0)\rangle \simeq \delta_{\alpha\alpha'}f_\alpha(x,t)\,,
\end{equation}
where initially $f_\alpha(x,0) = \delta(x)$, and the functions $f_\alpha$ satisfy the memory equation
\begin{equation}\label{2.5}
\partial_t f_\alpha(x,t)= \left(-c_\alpha \partial_x+D_\alpha \partial^2_x\right) f_\alpha (x,t) + \int^t_0 \int_{\mathbb{R}}
  f_\alpha(x-y,t-s) \partial^2_y M_{\alpha\alpha}(y,s)\,dy \, ds,
\end{equation}
$\alpha= 1,2$, where we have introduced $D_{\alpha\alpha}=D_\alpha$ and the memory kernel
\begin{equation}\label{2.6}
M_{\alpha\alpha}(x,t)=2 \sum_{\alpha',\alpha''=1,2} \left(G^\alpha_{\alpha'\alpha''}\right)^2 f_{\alpha'}(x,t) f_{\alpha''}(x,t)\,.
\nonumber
\end{equation}
If $\alpha' \neq \alpha''$, the product $f_{\alpha'}(x,t) f_{\alpha''}(x,t)$ is very small everywhere and hence can safely be neglected. Thereby the memory kernel  simplifies to 
\begin{equation}\label{2.6a}
M_{\alpha\alpha}(x,t)=2 \sum_{\alpha'=1,2} \left(G^\alpha_{\alpha'\alpha'}\right)^2 f_{\alpha'}(x,t)^2\,.
\end{equation}

To obtain the asymptotic behavior, one makes an educated scaling ansatz for $f_\alpha$, the precise 
computation being shifted to Appendix \ref{secA}. Particular cases were already presented in~\cite{Spohn14}. 
The universality classes are labeled according to whether
the leading coefficient $G^\alpha_{\alpha\alpha}$ vanishes or not. Each class still subdivides according to
the sub-leading terms  $G^{\alpha}_{\alpha'\alpha'}$. In our tables
$``1"$ indicates any value different from $0$, ``KPZ'' labels the scaling  reported in~(\ref{2.2}), ``$\alpha$-Levy'' 
a scaling determined by the maximally asymmetric $\alpha$-stable law with exponent $\alpha$, 
see~\eqref{eq:prediction_f2}  and~\eqref{eq:generic_Levy} below, and ``diff" a Gaussian peak with width proportional to $\sqrt{t}$.\medskip\\
Table 1:
\begin{center}
\begin{tabular}{l|cc|cc}
$G^1_{11}= 1$, $G^2_{22} = 1$  & $G^1_{22}$ & $G^2_{11}$ & peak 1 & peak 2\\[2pt]
\hline
& 0,1 & 0,1 & KPZ&KPZ 
\end{tabular}
\end{center}
In fact, the KPZ scaling function is not a solution of the fixed point equation derived from the  
mode-coupling equations (\ref{2.5}) - (\ref{2.6a}), but it turns out to be very close to this solution, 
see the discussion in~\cite{MS13}. 
\medskip\\Table 2:
\begin{center}
\begin{tabular}{l|cc|cc}
$G^1_{11}= 1$, $G^2_{22} = 0$  & $G^1_{22}$ & $G^2_{11}$ & peak 1 & peak 2\\[2pt]
\hline
& 0,1& 1& KPZ & $\tfrac{5}{3}$-Levy \\
& 1& 0& mod. KPZ & diff \\
& 0& 0& KPZ & diff 
\end{tabular}
\end{center}
As explained in more detail in Appendices \ref{secA} and \ref{sec:modified_KPZ},  if $G_{11}^2 \neq 0$, then the KPZ peak 1 
feeds into mode 2 to generate a peak 2 with $\tfrac{5}{3}$-Levy asymptotics, while the reverse process yields only a 
subdominant contribution to peak 1. On the other hand, if $G_{11}^2 =0 $ but $G_{22}^1 \neq 0$, then
on the level of mode-coupling the diffusive peak 2 generates a feedback on mode 1 which has also a 
dynamical exponent $z = 3/2$. In principle this should lead to a modified KPZ scaling function for peak 1.
If actually correct, the decoupling hypothesis would have to be slightly modified.
\medskip\\Table 3:
\begin{center} 
\begin{tabular}{l|cc|cc}
 $G^1_{11}= 0$, $G^2_{22} = 0$  & $G^1_{22}$ & $G^2_{11}$ & peak 1 & peak 2\\[2pt]
\hline
& 1& 1& \emph{gold}-Levy&\emph{gold}-Levy \\
& 1& 0& $\tfrac{3}{2}$-Levy & diff \\
& 0& 1& diff & $\tfrac{3}{2}$-Levy \\
& 0& 0& diff&diff 
\end{tabular}
\end{center}\bigskip
The case \textit{gold}-Levy is discussed in Appendix \ref{secA}.
If one peak is diffusive, it feeds back to the other peak, which then becomes $\tfrac{3}{2}$-Levy. 

The maximal asymmetry of the Levy distributions follows from the mode-coupling equations. But there is also a more qualitative argument. Physically one expects to have exponentially small correlations away from the sound cone 
$[c_1t,c_2t]$. If the Levy distribution would not be maximally asymmetric, then it would exhibit both-sided power law tails which necessarily  have slow decay outside the sound cone. Only for the maximal asymmetric distribution there is rapid decay to the outside and slow decay to the inside of the sound cone (see the discussion in Appendix~\ref{sec:properties_scaling_fcts}). In fact, in accordance with the general principle, in numerical simulations one always observes the Levy tail to be cut off at the other peak.

While we explained the asymptotic behavior of two coupled Burgers equation, one still has to relate them to a microscopic type model. In principle the theory should be applicable to any system with local interactions, either classical or quantum Hamiltonian, or classical with stochastic dynamics. Of course the model must have  exactly two conservation laws and the dynamics should be sufficiently chaotic so to have good space-time mixing properties. In all examples investigated in more detail the steady state can be written in product form. This has the advantage, that the Euler currents and $c_\alpha$ are known explicitly. After transformation to normal modes, the universality class for the model under consideration
can be easily determined. In fact, beyond the specific predictions, one strength of the theory is capture exceptional 
classes which would be hard to guess from a mere inspection of the equations of motion.  

Below we report on numerical solutions of a one-dimensional lattice theory, for which the field takes real values and is governed
by a deterministic differential equation plus random exchanges. We will present two examples for KPZ plus $\tfrac{5}{3}$-Levy peak, corresponding to table~2, row~1.
The same model with a harmonic interaction belongs to the class diffusive plus $\tfrac{3}{2}$-Levy peak (see Appendix~\ref{sec:expressions_G_specific_pot}). In this case a complete mathematical proof is available  \cite{BGJ14}, which validates the prediction from mode-coupling. Also stochastic lattice gas models with two species of particles have been investigated. In 
\cite{FeSS13} both peaks are KPZ. The two-lane model  \cite{PSS14} has more parameters. Generically the two peaks are KPZ, but also the universality class studied here can be realized. In the very recent contribution \cite{PSS14a} even more classes, including  \textit{gold}-Levy, are obtained. 
Finally we should mention the discrete non-linear Schr\"{o}dinger equation on a one-dimensional lattice with 
repulsive on-site interactions \cite{KuLa13,KuHuSp14,MeSp14b}. At low temperatures the model has the usual three conservation laws to a very good approximation. However the heat mode has a very small amplitude and one is reduced to an effective two-component system governing superfluid density and momentum. In this case both peaks are predicted to be KPZ, which is well confirmed through numerical simulations.

 %------------------------------------- hydrodynamics for KVM -------------------------------------
\section{The BS model with random exchanges}
\label{sec3}
\setcounter{equation}{0}

We consider the model as proposed and studied in~\cite{BS12}, called `BS' for short. Originally the model was motivated by anharmonic chains, for which stochastic collisions are added so to improve space-time mixing properties. One considers a real-valued 
field, denoted by $\eta_i\in \mathbb{R}$, $i \in \mathbb{Z}$. To define the model we first take a finite volume with $0 \leq i \leq N-1$. We call $\bm{\eta} = (\eta_0,\dots,\eta_{N-1})$ the displacement field, also `volume' and `height' have been proposed.
The dynamics of the BS model consists of a deterministic part, which describes forces exerted by neighboring displacements and 
a stochastic part in which neighboring displacements are exchanged at random. The deterministic part is governed by
the first order differential equations
\begin{equation}
  \label{eq:deterministic_dynamics}
  \frac{d}{dt} \eta_i= V'(\eta_{i+1})-V'(\eta_{i-1})
\end{equation}
and has the corresponding generator 
\[
\mathcal{A}_N = \sum_{i=0}^{N-1} \big( V'(\eta_{i+1})-V'(\eta_{i-1}) \big) \partial_{\eta_i}.
\]
Periodic boundary conditions are imposed as $\eta_{i+N} = \eta_i$. The potential $V$ is bounded from below with at least a one-sided growth to infinity as $|\eta_i |\to \infty$.
In addition, at independent random times distributed according to an exponential law with parameter~$\gamma$, neighboring displacements are exchanged. The generator for the random part is $\gamma \mathcal{S}_N$ with
\[
\mathcal{S}_Nf(\bm{\eta}) = \sum_{i=0}^{N-1}\big( f\left(\bm{\eta}^{i,i+1}\right) - f(\bm{\eta})\big), 
\qquad 
\bm{\eta}^{i,i+1} = \left(\eta_0,\dots,\eta_{i-1},\eta_{i+1},\eta_i,\eta_{i+2},\dots,\eta_{N-1}\right).
\]

Clearly the displacement field is locally conserved. Note that under the deterministic part
\begin{equation}
  \label{eq:energy}
  \frac{d}{dt} V(\eta_i) = V'(\eta_{i+1})V'(\eta_{i})-V'(\eta_{i})V'(\eta_{i-1}).
\end{equation}
Thus, including random exchanges, also $V(\eta_i)$ is locally conserved. This field is called the (potential) energy field.
As a consequence, the BS model has  a two-parameter family of invariant measures. The parameter dual to  
$V(\eta_i)$ is called inverse temperature, denoted by $\beta >0$, and the parameter dual to $\eta_i$ is called tension, denoted by $\tau \in \mathbb{R}$. Hence the invariant measures are written as
\begin{equation}
\label{eq:grand_canonical_measure}
\mu_{\tau,\beta}\left(d\eta_0\dots d\eta_{N-1}\right) = \prod_{i=0}^{N-1} Z_{\tau,\beta}^{-1} \, \rme^{-\beta(V(\eta_i)+\tau \eta_i)} \, d\eta_i.
\end{equation}
If the potential increases too slowly as $\eta_i \to \pm \infty$, the range of admissible values of $\tau$ may have to be restricted in order for the density $\rme^{-\beta(V(\eta)+\tau \eta)}$ to be integrable. 
(We use $\eta \in \mathbb{R}$ as standing for one of the $\eta_i$'s.)
Averages with respect to~$\mu_{\tau,\beta}$ are denoted by $\langle\cdot\rangle_{\tau,\beta}$. 

At finite volume the micro-canonical measures are time-invariant, but they could be not ergodic. Such possible pathology disappears in the infinite volume limit. In~\cite{BS12} it is established that the infinite volume dynamics is ergodic, in the sense that all invariant measures of the dynamics of finite relative entropy with respect to the infinite dimensional analogue of~$\mu_{0,1}$ and which are translation invariant, are convex combinations of canonical measures. Hence, in the infinite volume limit,  displacement and energy are the only conserved fields.

The local conservation of displacement and energy implies the existence of local displacement and energy currents. They have a deterministic and random part with the former given by 
\begin{equation}
\label{eq:current}
\frac{d}{dt}\begin{pmatrix} \eta_i \\ V(\eta_i) \end{pmatrix} = J^{i-1,i} - J^{i,i+1}, \qquad J^{i,i+1} = \begin{pmatrix} j_h^{i,i+1} \\ j_e^{i,i+1} \end{pmatrix} = -\begin{pmatrix} V'(\eta_i) + V'(\eta_{i+1}) \\ V'(\eta_i)V'(\eta_{i+1}) \end{pmatrix}.
\end{equation}

To apply the theory from Section \ref{sec2}, we first have to obtain the macroscopic Euler equations. In the continuum limit, studied in~\cite{BS12}, the 
displacement field becomes $h(x,t)$ and the energy field  $e(x,t)$. The currents of the Euler equations are determined by averaging the currents in a local equilibrium state. On that scale the random exchange makes no contribution  yet and it suffices to compute the average of the currents in (\ref{eq:current}) with respect to $\mu_{\tau,\beta}$. Since
\[
\langle V'(\eta_i) \rangle_{\tau,\beta} = -\tau, 
\qquad 
\langle V'(\eta_i) V'(\eta_{i+1}) \rangle_{\tau,\beta} = \tau^2,
\]
the Euler currents for the conserved fields $h,e$ are respectively $j_h = 2 \tau$ and $j_e = -\tau^2$, where the tension
$\tau$ is considered as a function of the average displacement and energy as
defined through the implicit relation
\begin{equation}
  \label{eq:impllcit}
h_{\tau,\beta} = \langle \eta_i \rangle_{\tau,\beta},
\qquad
e_{\tau,\beta} = \langle V(\eta_i) \rangle_{\tau,\beta}.
\end{equation}
In the hydrodynamic limit the system of conservation laws then reads
\begin{equation}
  \label{eq:Euler}
\partial_t \begin{pmatrix} h(x,t) \\ e(x,t) \end{pmatrix} + \partial_x \begin{pmatrix} 2 \tau(h(x,t),e(x,t)) \\[7pt] -\tau(h(x,t),e(x,t))^2 \end{pmatrix} = 0.
\end{equation}
The linearization of this system around a uniform background profile $(h_0,e_0)$, obtained by writing $h(x,t) = h_0 + \tilde{h}(x,t)$ and $e(x,t) = e_0 + \tilde{e}(x,t)$, yields
\begin{equation}
  \label{eq:linearization}
  \partial_t \!\begin{pmatrix} \tilde{h}(x,t) \\ \tilde{e}(x,t) \end{pmatrix} + A(h_0,e_0) \partial_x \!\begin{pmatrix} \tilde{h}(x,t) \\ \tilde{e}(x,t) \end{pmatrix} = 0,
\end{equation}
where
\begin{equation}
  \label{eq:linearization1}
A = 2 \begin{pmatrix}
\partial_h \tau & \partial_e \tau \\
-\tau \partial_h \tau & -\tau \partial_e \tau
\end{pmatrix}.
\end{equation}
In (\ref{eq:linearization1})  the dependence of $A$ on  $h_0,e_0$ has already been suppressed. In the sequel we will regard $\tau,\beta$ as given and thereby via ({\ref{eq:impllcit}) also the value of the background fields $h_0,e_0$.

We now follow the strategy in~\cite{Spohn14} in order to study the space-time correlation matrix $S(i,t) \in \mathbb{R}^{2 \times 2}$ of the conserved fields whose entries read
\[
S_{\alpha\alpha'}(i,t) = \left\langle g_\alpha(\eta_{i,t})  g_{\alpha'}(\eta_{0,0})\right\rangle_{\tau,\beta} - \left\langle g_\alpha(\eta_{i,t}) \right\rangle_{\tau,\beta} \left\langle g_{\alpha'}(\eta_{0,0})\right\rangle_{\tau,\beta}.
\]
Here $g_1(\eta) = \eta$ and $g_2(\eta) = V(\eta)$ and, in slight abuse, $\langle\cdot\rangle_{\tau,\beta}$ refers to average in the stationary process with starting measure $\mu_{\tau,\beta}$.
For the said purpose we expand the Euler equations (\ref{eq:Euler}) to second order in the currents and add dissipation plus noise. The resulting Langevin equations have a structure similar to (\ref{2.3}), but with the linear drift term not yet diagonal. The latter feature is accomplished through the transformation matrix $R$ defined by  the properties
\[
R A R^{-1} = \mathrm{diag}(c,0),\qquad RS(0,0)R^\mathrm{T} = 1,
\]
where it is  already anticipated that $A$ has the eigenvalues $0$ and
 \begin{equation}
  \label{eq:sound_speed}
  c = 2 (\partial_h  -\tau \partial_e) \tau < 0,
\end{equation}
see~\eqref{eq:def_c}. We use the convention that the left moving mode has label $1$, while the standing mode has label $2$. In analogy to anharmonic chains,  mode $1$ is called sound mode and mode $2$ heat mode. After this transformation the equations of nonlinear fluctuating hydrodynamics are exactly of the form of two coupled Burgers equations as in~(\ref{2.3}). 
The transformation matrix $R$ and the nonlinear coupling matrices $G^{\alpha}$ are tabulated in~\eqref{eq:matrix_R} and Appendix~\ref{sec:G_matrices} respectively. Because of the particular form of the Euler currents, one has  $G^2_{22} = 0$, $G^2_{12} = G^2_{21} =0$ always, while 
$G^2_{11} < 0$. Thus the heat peak is non-KPZ, but coupled to the sound peak. According to our classification, this leaves only the two cases: \textit{(i)} $G^1_{11} \neq 0$ implying KPZ for mode~$1$ and $\tfrac{5}{3}$-Levy for mode~$2$,  \textit{(ii)} $G^1_{11} = 0$ implying diffusive for mode~$1$ and $\tfrac{3}{2}$-Levy for mode $2$. 
  
The case \textit{(ii)} is exceptional, a more explicit condition being 
\[ 
(\partial_h - \tau \partial_e)^2\tau = 0,
\]  
see~\eqref{eq:G^1_11}. One example is the harmonic potential $V(\eta) = \eta^2$ discussed in \cite{BGJ14}, for which $G^1_{11} = 0$
identically. In general, there could be special values of  $\tau,\beta$ at which  $G^1_{11} = 0$.

The matrix $S(i,t)$ is transformed to normal modes  as $S^\sharp(i,t) = R S(i,t) R^\mathrm{T}$.
On sufficiently large scales $ S^\sharp(i,t)$ should be determined through the stationary covariance of the coupled Burgers equations (\ref{2.3})
and we can use directly the results from Section \ref{sec2} and Appendix \ref{secA}. They assert that $R S(i,t) R^\mathrm{T}$
is approximately diagonal,
\begin{equation}
  \label{eq:normal_mode_transform}
  S^\sharp(i,t) = R S(i,t) R^\mathrm{T} \simeq \delta_{\alpha\alpha'} \, f_\alpha(N^{-1}i,t), \quad \mathrm{mod}\,\,N \, .
\end{equation}
The sound peak scales asymptotically as
\begin{equation}
\label{eq:prediction_f1}
f_1(x,t) \simeq (\lambda_1 t)^{-2/3} f_{\rm KPZ}((\lambda_1 t)^{-2/3}(x-ct)), 
\qquad 
\lambda_1 = 2\sqrt{2} \left|G_{11}^1 \right|,
\end{equation}
and the heat peak as 
\begin{equation}
\label{eq:prediction_f2}
f_2(x,t) \simeq (\lambda_2 t)^{-3/5} f_{{\rm Levy},5/3,1}((\lambda_2 t)^{-3/5}x), 
\qquad 
\lambda_2 = a_{\rm h} \, c^{-1/3} \left( G^2_{11} \right)^2 \lambda_1^{-2/3}, 
\end{equation}
with the constant\footnote{Anharmonic chains evolving according to Hamiltonian dynamics
have three conservation laws and correspondingly one heat mode and two reflection symmetric sound modes.
The heat mode is the symmetric $\tfrac{5}{3}$-Levy function and the prefactor is $2 a_{\rm h}$ 
(compare for instance with~\cite[Equation~(50)]{MS14}).}
\[
a_{\rm h} = \sqrt{3} \, \Gamma\big(\tfrac13\big) \int_\mathbb{R} \left(f_{\rm KPZ}\right)^2 \simeq 1.81.
\]

%---------------------------- Numerical results ---------------------
\section{Numerical simulations for the BS model}
\label{sec:numerics}

To assess the validity of~\eqref{eq:prediction_f1} - \eqref{eq:prediction_f2}, we perform numerical simulations obtained by integrating the dynamics~\eqref{eq:deterministic_dynamics} and adding random exchanges when starting from initial conditions distributed according to the canonical measure~\eqref{eq:grand_canonical_measure}. This is done for two potentials: the FPU-$\alpha$ potential (simply abbreviated as FPU in the sequel)
\begin{equation}
\label{eq:FPU_pot}
V(\eta) = \frac12 \eta^2 + \frac{a}{3}\eta^3 + \frac14 \eta^4,
\end{equation}
with $a = 2$ (as in~\cite{MS13}), and the Kac-Van Moerbeke potential (abbreviated as KvM in the sequel)
\begin{equation}
\label{eq:KVM_pot}
V(\eta) = \frac{\rme^{-\kappa \eta}+ \kappa\eta -1}{\kappa^2},
\end{equation}
with $\kappa=1$. Note that $\tau > -1/\kappa$ is necessary to ensure the normalization of the canonical measures in this case. The KvM potential is special since it makes the system integrable in the absence of stochastic exchanges, \textit{i.e.} when $\gamma = 0$, see~\cite{KvM75}. In fact, the corresponding system is related by a simple transformation to the famous Toda lattice~\cite{Toda88}, \emph{i.e.} a chain of oscillators coupled through the potential \eqref{eq:KVM_pot} to nearest neighbors and evolving according to Hamiltonian dynamics. 

In Section~\ref{sec:computation_correlators} we explain how the correlators are computed and fitted to the respective scaling functions. In Section~\ref{sec:comparison} we quantify the agreement between the numerically computed correlators and the theoretical predictions.

\subsection{Numerical computation of the correlation functions}
\label{sec:computation_correlators}

\subsubsection{Generation of initial conditions}

Initial conditions are sampled according to the canonical measure~\eqref{eq:grand_canonical_measure}. Since this measure is of product form, one can sample independently the initial values $(\eta_{i,0})_{i=0,\dots,N-1}$ for each site according to the measure $Z_{\tau,\beta}^{-1} \, \rme^{-\beta(V(\eta)+\tau \eta)} \, d\eta$. One way to do so is to start from $\eta^{\rm init}_i = 0$ and to evolve according to the SDE
\[
d\eta_{i,t}^{\rm init} = - \big(V'(\eta_{i,t}^{\rm init}) + \tau \big)dt + \sqrt{\frac{2}{\beta}} \, dW_{i,t}^{\rm init}.
\]
In practice this is done  by a discretization through an Euler-Maruyama scheme, using a time step $\Delta t_{\rm thm}$,
\[
\eta_{i,n+1}^{{\rm init}} = \eta_{i,n}^{{\rm init}} - \Delta t_{\rm thm} \Big( V'(\eta_{i,n}^{{\rm init}}) + \tau\Big) + \sqrt{\frac{2 \Delta t_{\rm thm}}{\beta}} \, G_{i,n}^{{\rm init}},
\]
where $G_{i,n}^{{\rm init}}$ are independent and identically distributed (i.i.d.) standard Gaussian random variables. We use $\Delta t_{\rm thm} = 0.005$, and integrate over $N_{\rm thm} = 1000$ steps. We finally set $\eta_{i,0} = \eta_{i,N_{\rm thm}}^{{\rm init}}$.
To check the sampling of the initial conditions, we compared the reference distribution $Z_{\tau,\beta}^{-1} \, \rme^{-\beta (V(\eta)+\tau \eta)}$ and the histogram of the displacements $(\eta_{i,N_{\rm thm}}^{{\rm init}})_{0 \leq i \leq N-1}$. 

\subsubsection{Numerical integration of the dynamics}

The dynamics~\eqref{eq:deterministic_dynamics} is integrated with a timestep $\Delta t > 0$ using the algorithm presented in~\cite[Section~6.3.1]{BS12}, adapted here to the periodic setting. In a nutshell, the numerical method first integrates the deterministic part of the dynamics over a time increment~$\Delta t$ with a splitting strategy where even and odd sites are evolved separately over time increments~$\Delta t$, in accordance with the hidden Hamiltonian structure of the deterministic part of the dynamics. Next one has to include the random exchanges: independent exponential clocks are attached to each pair $(\eta_i,\eta_{i+1})$, and the current clock times  are decreased by~$\Delta t$ at each time step. When a clock time becomes negative, the corresponding neighboring displacements are exchanged, and a new exponential time of mean~$1/\gamma$ is sampled.

We produce $K$ samples of initial conditions of the system (starting from independent initial conditions $\eta_{i,0}^k$), and denote by $(\eta_{i,n}^{k})_{0 \leq i \leq N-1}$ an approximation of the state of the $k$th sample at time $n\Delta t$. The time step is set to $\Delta t = 0.005$, a value sufficiently small to ensure relative energy variations of order $10^{-3}$ or less over very long times for stochastic rates in the range $0 \leq \gamma \leq 1$ and for system sizes up to $N = 8000$. Note that the splitting algorithm respects the underlying symplectic structure of the differential equation part, while the exchange does not. As a consequence the near energy conservation observed for the deterministic dynamics is degraded by the exchange noise, although no systematic drift is observed.

\subsubsection{Computation of the correlators}

The correlation matrices are computed at times~$n\Delta t$. To this end, we first evaluate the empirical average over the replicas of the displacements and energies at each site~$i$,
\[
\overline{h}_{i,n} = \frac1K \sum_{k=1}^K \eta_{i,n}^{k}, \qquad \overline{e}_{i,n} = \frac1K \sum_{k=1}^K V\left(\eta_{i,n}^{k}\right),
\]
and then compute the entries of the correlation matrix by the following space- 
and sample-average,
\[
\left[ C_{N,K}(i,n) \right]_{\alpha,\alpha'} = \frac{1}{NK} \sum_{k=1}^K \sum_{i=0}^{N-1} u_{\alpha,i+j,n}^{k} u_{\alpha',j,0}^{k},
\]
with 
\[
u_{1,i,m}^{k} = \eta_{i,m}^{k} - \overline{h}_{i,m}, \qquad u_{2,i,m}^{k} = V\left(\eta_{i,m}^{k}\right) - \overline{e}_{i,m}.
\]
The numerical results reported below are obtained using $K = 10^5$ samples and chains of lengths $N=2000-8000$. We checked that
\[
C_{N,K}(i,0) \simeq \delta_{i0} \, \mathcal{C}, \qquad \mathcal{C} = \begin{pmatrix} \langle \eta;\eta\rangle_{\tau,\beta}  & \langle \eta;V(\eta)\rangle_{\tau,\beta} \\ \langle \eta;V(\eta)\rangle_{\tau,\beta} & \langle V(\eta);V(\eta)\rangle_{\tau,\beta} \end{pmatrix},
\]
where $\left \langle A;B\right\rangle_{\tau,\beta} = \left \langle A(\eta_0) B(\eta_0) \right\rangle_{\tau,\beta} - \left \langle A(\eta_0) \right\rangle_{\tau,\beta}\left \langle B(\eta_0) \right\rangle_{\tau,\beta}$. We also checked that the sum rules hold, up to very small errors related to the only approximate conservation of the energies and to the finiteness of the number of samples~$K$,
\[
\sum_{i=0}^{N-1} C_{N,K}(i,n) = \sum_{i=0}^{N-1} C_{N,K}(i,0).
\]
After a normal mode transformation as in~\eqref{eq:normal_mode_transform}, based on the matrix~$R$ defined in~\eqref{eq:matrix_R}, the correlation matrix is almost diagonal,
\[
C^\sharp_{N,K}(i,n) = R C_{N,K}(i,n) R^\mathrm{T} \simeq \begin{pmatrix} f_1^{\rm num}(i,n) & 0 \\ 0 & f_2^{\rm num}(i,n) \end{pmatrix}.
\]

\subsubsection{Computation of the scaling factors}
\label{sec:optimization}

% execute 'complete_optimization.sce' and then 'exponent.sce' to perform the fit
In order to check quantitatively the agreement between the numerically computed correlation functions $f_\alpha^{\rm num}$ and the theoretically predicted values~\eqref{eq:prediction_f1}--\eqref{eq:prediction_f2},  following~\cite{MS14},
we optimize the parameters in the ansatz~\eqref{eq:prediction_f1}--\eqref{eq:prediction_f2} such as to minimize the $L^1$ distance, 
\begin{equation}
\label{eq:minimization_pbm}
\inf_{\substack{x_n \in \mathbb{R} \\ \Lambda_n > 0}} \left\{ \sum_{i=0}^{N-1} \left| f_\alpha^{\rm num}(i,n) - (\Lambda_n)^{-1} f_\alpha^{\rm mc}\big((\Lambda_n)^{-1} (i-x_n)\big)\right| \right\}. 
\end{equation}
Here, $f_\alpha^{\rm mc}$ denotes the theoretical scaling function, namely KPZ for mode~1 and maximally asymmetric $\tfrac{5}{3}$-Levy for peak~2. In fact, in order to have a more stable minimization procedure, we use the prior knowledge on how $x_n,\Lambda_n$ should scale and write
\begin{equation}
\label{eq:guessed_scalings}
x_n = c_{\rm theor} n\Delta t + \widetilde{x}_n, \qquad \Lambda_n = \widetilde{\Lambda}_n \, \left(n \Delta t\right)^{\delta_{\rm theor}}.
\end{equation}
The value $c_{\rm theor}$ is the theoretical peak velocity, to say, 0 for the heat peak and~\eqref{eq:sound_speed} for the sound peak, and $\delta_{\rm theor}$ the theoretically predicted scaling exponent, 3/5 for the heat mode and 2/3 for the sound peak. 

The optimization in~\eqref{eq:minimization_pbm} is now performed over $\widetilde{x}_n$ and $\widetilde{\Lambda}_n$ at the various times $n\Delta t$ at which the correlation matrix is computed. In practice, the sum in~\eqref{eq:minimization_pbm} is not performed over all indices~$i$ but restricted to the indices~$i$ which are close to the center of the peak under investigation, since far away from the peak center the correlation is almost zero and the dominance of statistical noise makes those values irrelevant. The center of the peak at time index~$n$ is defined as the index $i^n_{\rm center}$ for which $f_\alpha^{\rm num}(i,n)$, as a function of~$i$, is maximal. A cut-off range $R_{\rm cut} > 0$ is then introduced to limit the sum in~\eqref{eq:minimization_pbm} to indices $i^n_{\rm center} - R_{\rm cut} t^{\delta_{\rm theor}} \leq i \leq i^n_{\rm center} + R_{\rm cut} t^{\delta_{\rm theor}}$.

The values $\widetilde{x}_n,\widetilde{\Lambda}_n$ may be drifting in time when the expected scaling~\eqref{eq:guessed_scalings} is not completely exact. It may happen for instance that, due to errors related to the use of finite stepsizes $\Delta t$, the actual velocity is not exactly equal to $c_{\rm theor}$. We therefore fit $\widetilde{x}_n,\widetilde{\Lambda}_n$ as
\begin{equation}
\label{eq:fit_tilde}
\widetilde{x}_n = c_{\rm crt} \, n\Delta t + x_0, \qquad \widetilde{\Lambda}_n = \widetilde{\Lambda}_0 \, (n\Delta t)^{\delta_{\rm crt}},
\end{equation}
these fits being performed using a standard least-square minimization for $\widetilde{x}_n$ and a least-square minimization based on $\log \widetilde{\Lambda}_n$ to find the correction to the scaling exponent. The actual velocity observed in the numerical experiments is then $c_{\rm num} = c_{\rm theor}+c_{\rm crt}$ and the actual scaling exponent is $\delta_{\rm num} = \delta_{\rm theor}+\delta_{\rm crt}$. Once the corrected scaling exponent is determined, the scaling factor is obtained as
\begin{equation}
\label{eq:fitted_scaling_factor}
\lambda_{\rm num} = \widetilde{\Lambda}_0^{1/\delta_{\rm num}}.
\end{equation}
We have checked that the final outputs, in particular the actual scaling exponent $\delta_{\rm num}$ and the associated scaling factor $\lambda_{\rm num}$ are insensitive to the choice of the surrogate scaling exponent $\delta_{\rm theor}$. This procedure also allows to check how fast the ``instantaneous'' estimates of the scaling factor, defined as
\begin{equation}
\label{eq:instantaneous_scaling_factor}
\lambda_n = \left( \widetilde{\Lambda}_n (n\Delta t)^{-\delta_{\rm crt}} \right)^{1/\delta_{\rm num}},
\end{equation}
stabilize around the average value~$\lambda_{\rm num}$ given by~\eqref{eq:fitted_scaling_factor}, see Figures~\ref{fig:cv_FPU_sound}, \ref{fig:cv_FPU_heat}, \ref{fig:cv_KVM_sound}, and~\ref{fig:cv_KVM_heat}, Left.

\subsection{Comparison with theoretical predictions}
\label{sec:comparison}

The numerical results reported here have been obtained at the fairly low temperature of $\beta ^{-1}= \tfrac{1}{2}$, with a tension $\tau = 1$, using a noise intensity $\gamma = 1$, and the value $R_{\rm cut} = 9$ for the sound peak and $R_{\rm cut} = 11$ for the heat peak to compute the $L^1$ error in the minimization procedure~\eqref{eq:minimization_pbm}. A plot summarizing the evolution of the sound and heat peaks is presented in Figure~\ref{fig:superposition}. In all cases, the value $x_0$ in~\eqref{eq:fit_tilde} is very small and is henceforth set to~0. 

In the pictures, we call ``rescaled peak'' the plots for which the renormalized numerical correlation functions $(\lambda_{\rm num}n\Delta t)^{\delta_{\rm num}} f_\alpha^{\rm num}(i,n)$ are plotted, at a given time index~$n$, as a function of the renormalized spatial variable $(i-c_{\rm num}n\Delta t)/(\lambda_{\rm num}n\Delta t)^{\delta_{\rm num}}$.

%obtained with GNUPLOT_SUPERPOSITION
\begin{figure}[h]
\includegraphics[width=15cm]{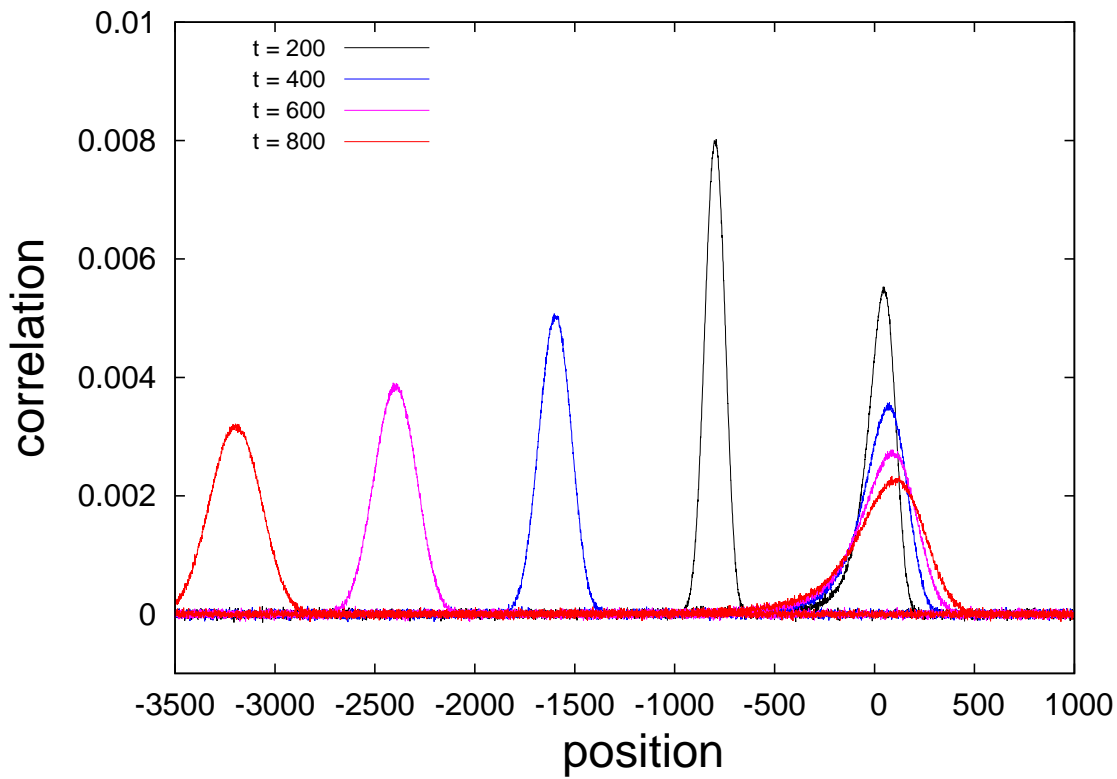}
\caption{\label{fig:superposition} Evolution of the heat peak (centered at~$x=0$) and the sound peak, traveling to the left, for the KvM potential~\eqref{eq:KVM_pot}. Note that the heat peak is not symmetric, the rapid decay being away from the sound peak.}
\end{figure}

%--------------- FPU -------------------
\subsubsection{FPU potential}

The rescaled sound and heat peaks for the FPU potential are presented in Figure~\ref{fig:comparison_FPU}. The agreement with the predicted scaling functions is qualitatively excellent. On a quantitative level, the numerical parameters obtained by the minimization procedure are:
\begin{itemize}
\item for sound peaks, exponent $\delta_{\rm num} = 2/3$, fixed to its theoretical value since $\delta_{\rm num}$ turns out to be extremely close to 2/3, velocity $c_{\rm num} = -5.24$, compared to the theoretical value $c_{\rm theor} = -5.28$, and scaling factor $\lambda_{1} \simeq 6.36$. The scaling factor is in excellent agreement with the theoretical value $\lambda_{1} = 2\sqrt{2} |G^1_{11}| = 6.32$ predicted by~\eqref{eq:prediction_f1}. 
\item for heat peaks, the reference being the maximally asymmetric Levy distribution with $\alpha = 5/3$: velocity $c_{\rm num} = 0$, exponent $\delta_{\rm num} = 0.605$, very close to the theoretical value $3/5$, scaling factor $\lambda_2 \simeq 3.70$. The scaling factor is in very good agreement with the theoretical value $3.46$ predicted by~\eqref{eq:prediction_f2}. A slightly better agreement could be obtained by decreasing a little bit the parameter of the Levy distribution from $5/3$ to values around to~1.64 in order to have a sharper decrease on the right.
\end{itemize}
The evolution of the non-universal scaling factors as a function of the time index is reported in Figures~\ref{fig:cv_FPU_sound} and~\ref{fig:cv_FPU_heat}, together with the $L^1$ error. Note that the error very quickly decreases at the beginning of the simulation but, after reaching an absolute minimum, slowly increases again due to the increase of the statistical noise. The initial decrease is faster for the sound peak, which attains  its asymptotic shape more rapidly. Also, the scaling factor settles down slightly faster for the sound peak. 

%obtained with COMPARE_GNUPLOT_SOUND_FPU
%obtained with COMPARE_GNUPLOT_HEAT_FPU
\begin{figure}[h]
\begin{center}
\includegraphics[width=7.6cm]{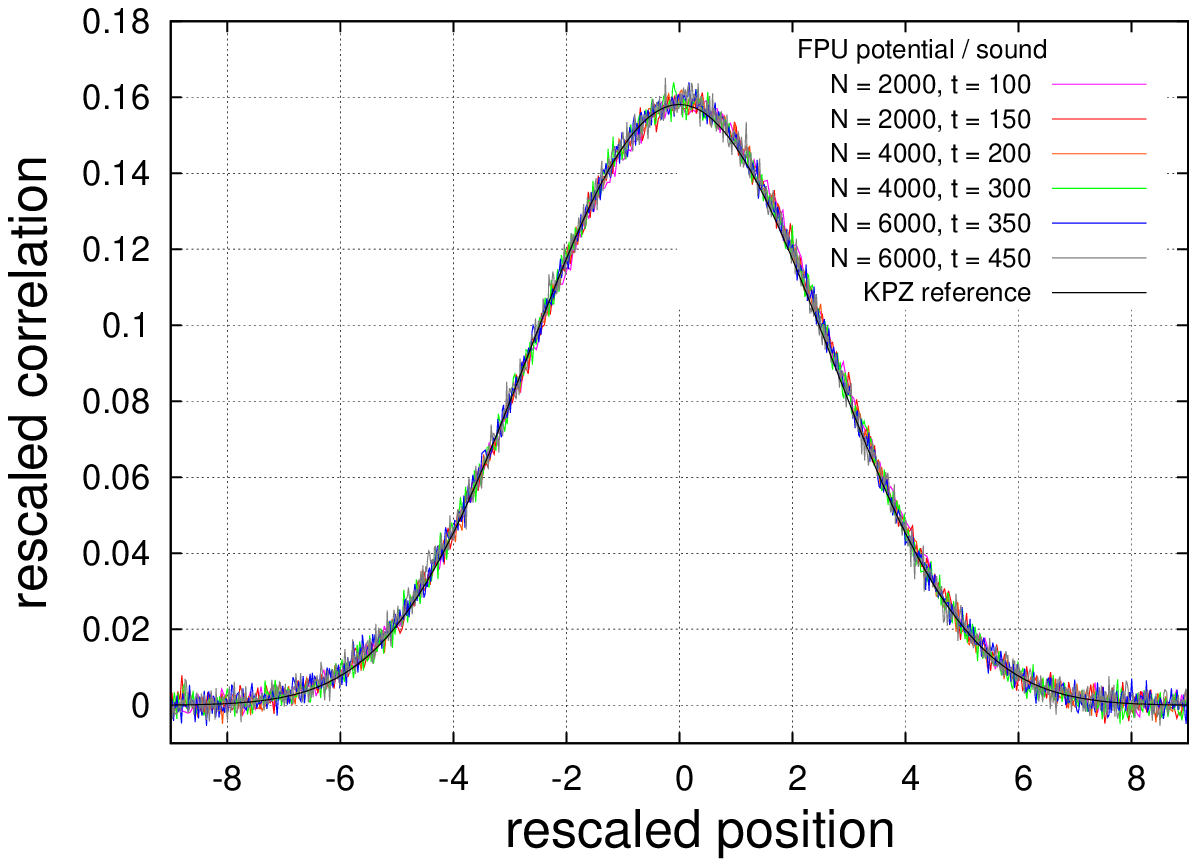}
\includegraphics[width=7.6cm]{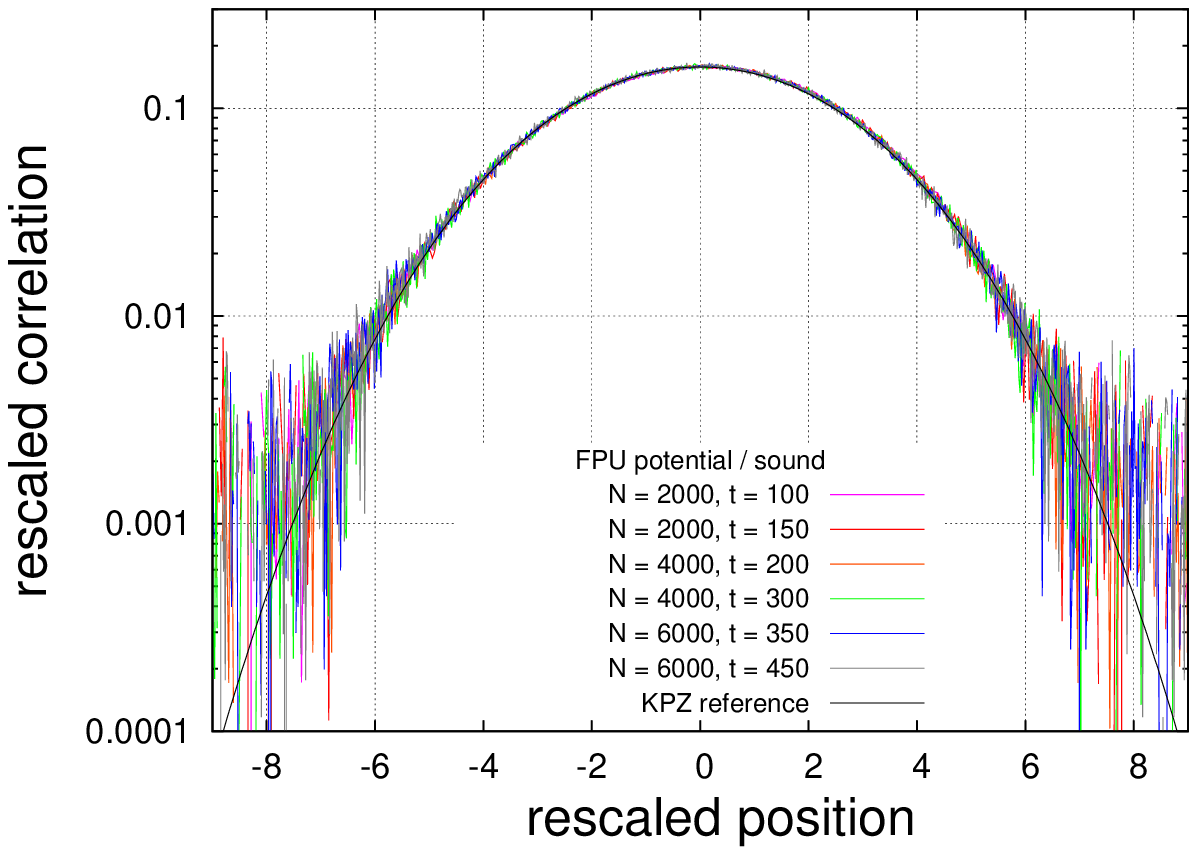} \\
\includegraphics[width=7.6cm]{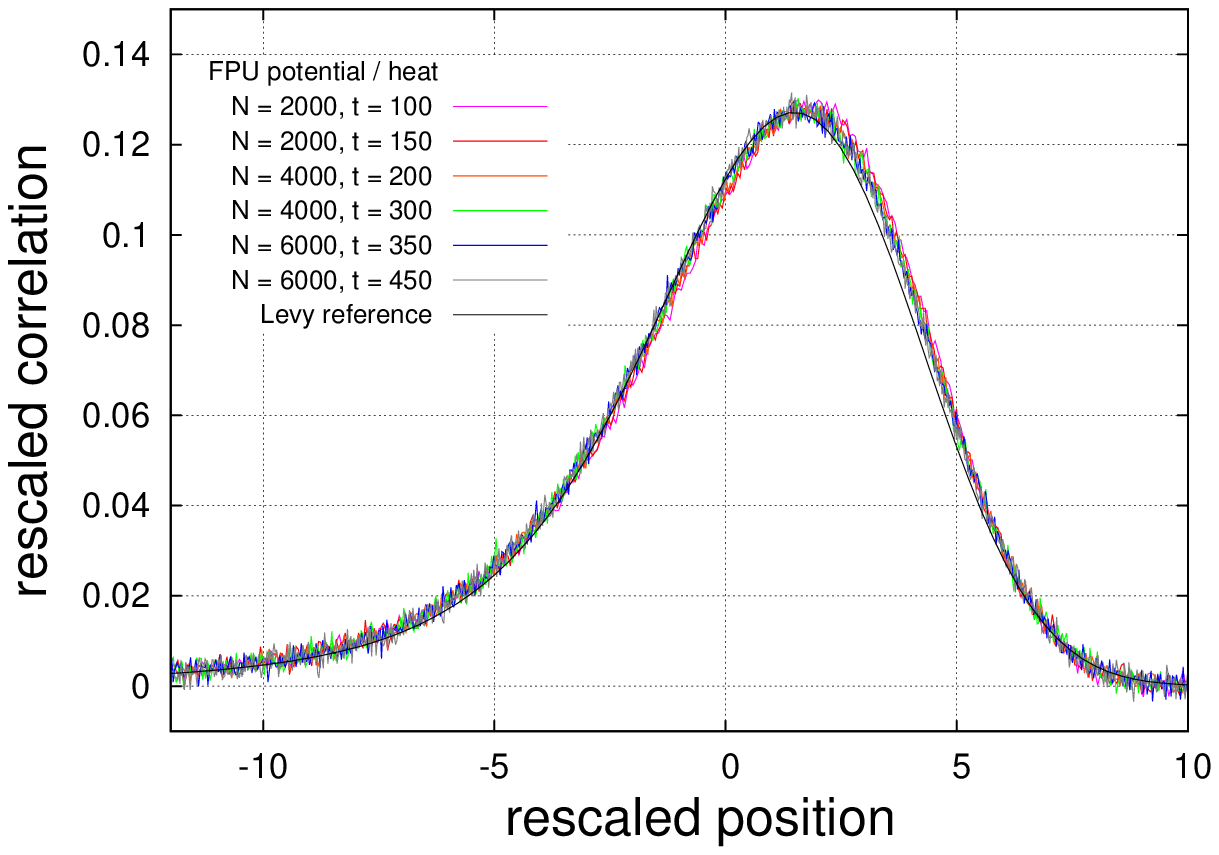}
\includegraphics[width=7.6cm]{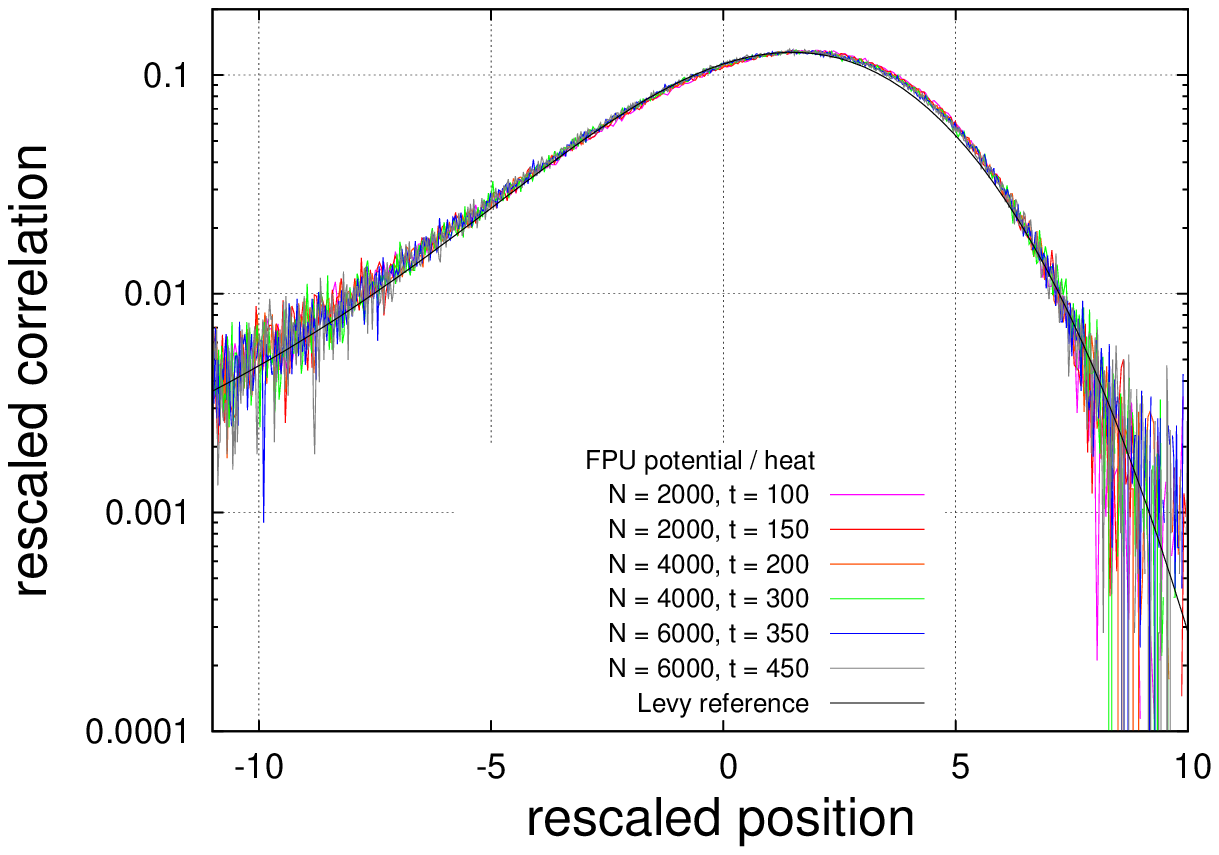}
\end{center}
\caption{\label{fig:comparison_FPU} (FPU potential) Comparison of rescaled sound and heat peaks. The first line corresponds to sound modes, the second to heat modes. The reference for heat modes is the Levy stable distribution with parameter $\alpha=5/3$ and maximal asymmetry. Logarithmic plots are provided in the right column.}
\end{figure}

\begin{figure}[h]
\includegraphics[width=7.5cm]{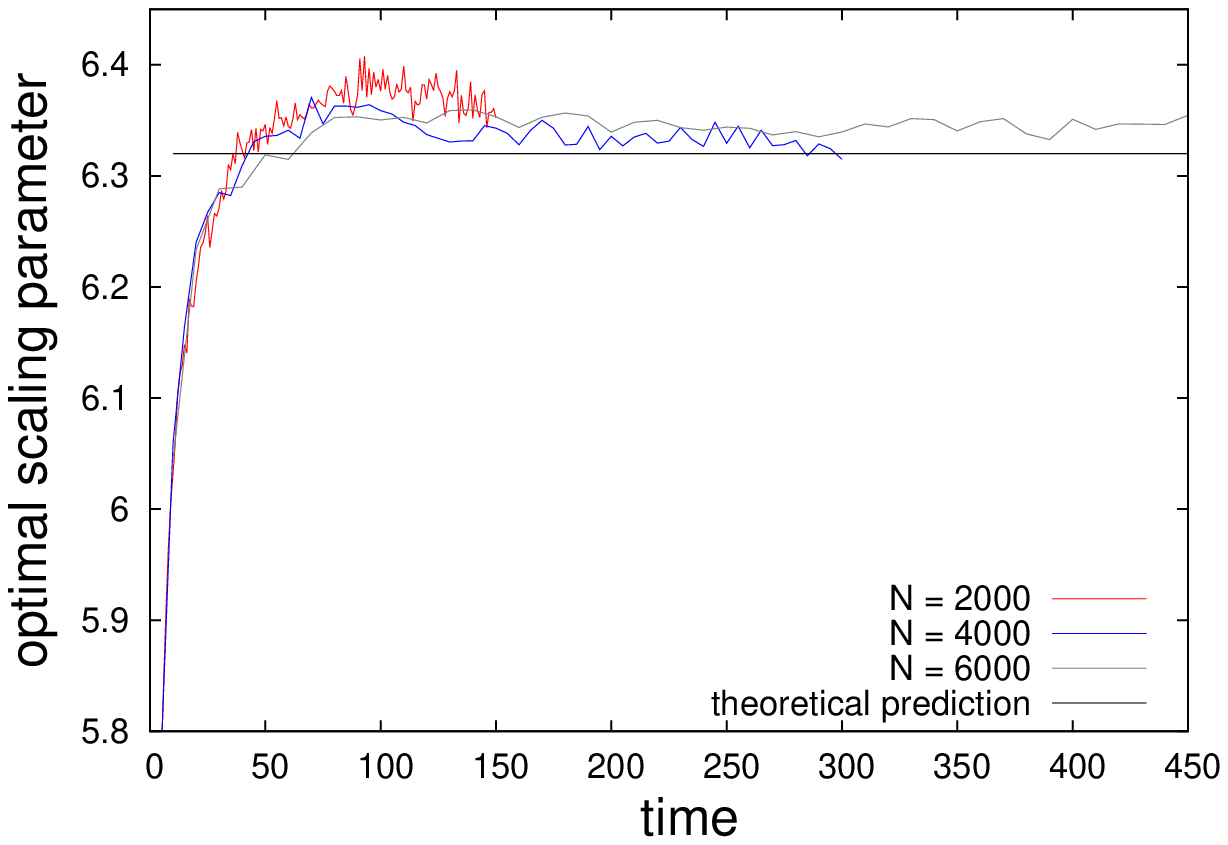}
\includegraphics[width=7.5cm]{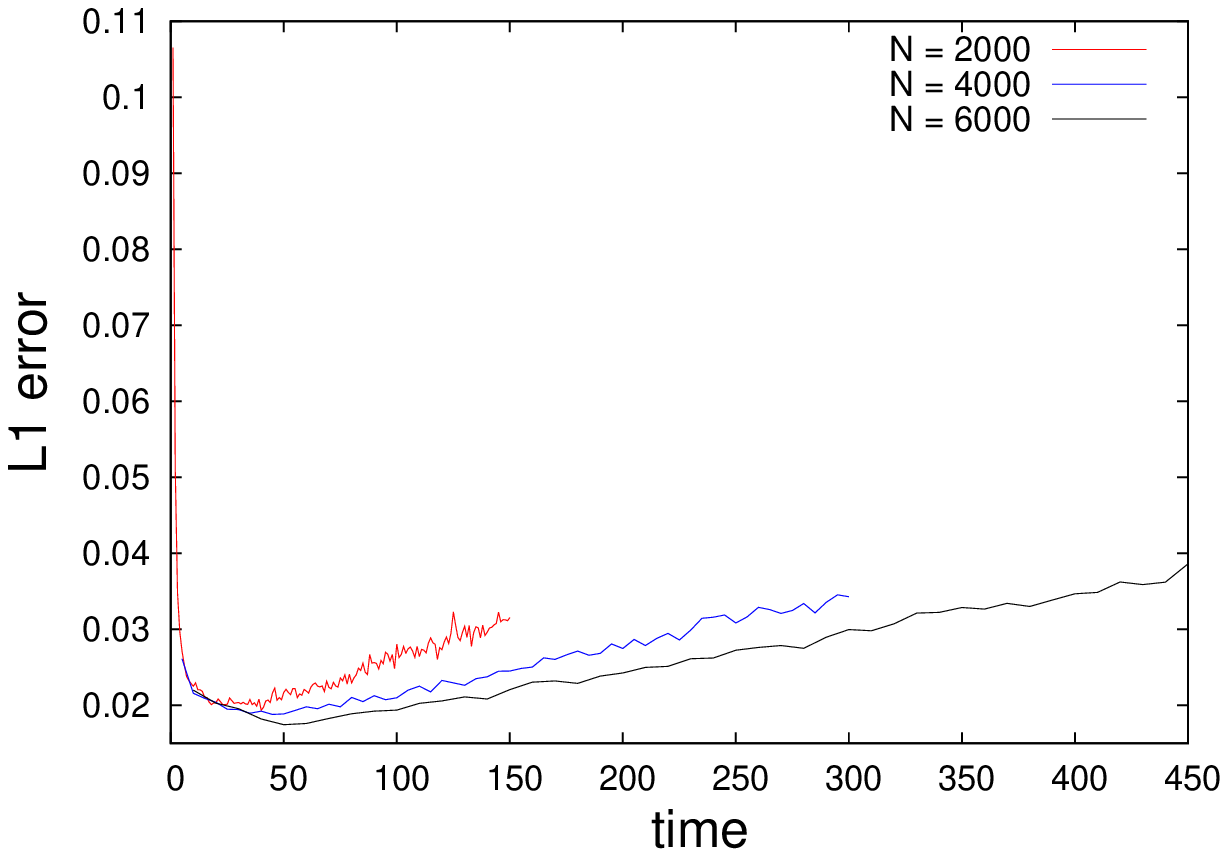}
\caption{\label{fig:cv_FPU_sound} (FPU potential, sound peak) Left: Optimal value of the scaling parameter for a given time, as given by~\eqref{eq:instantaneous_scaling_factor}. Right: $L^1$ error~\eqref{eq:minimization_pbm} for the optimal value of the parameters.}
\end{figure}

\begin{figure}[h]
\includegraphics[width=7.5cm]{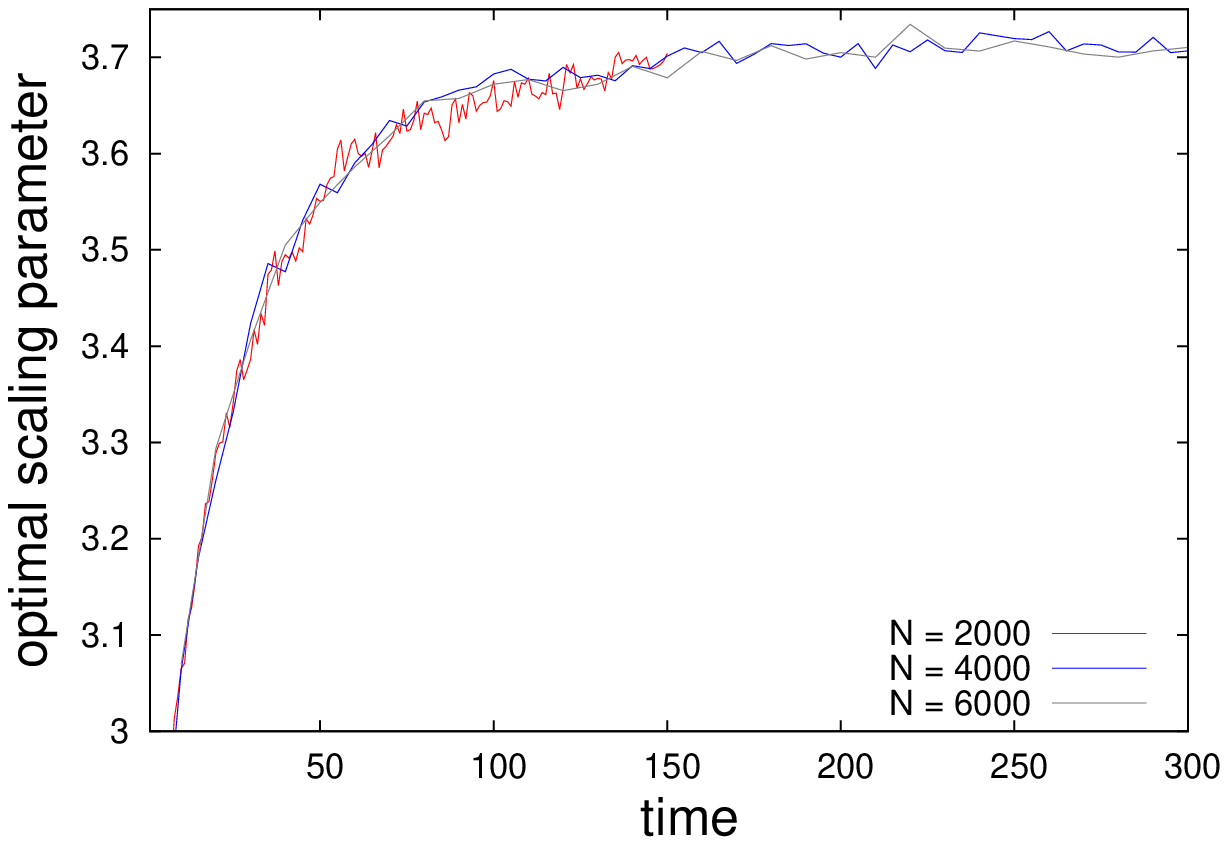}
\includegraphics[width=7.5cm]{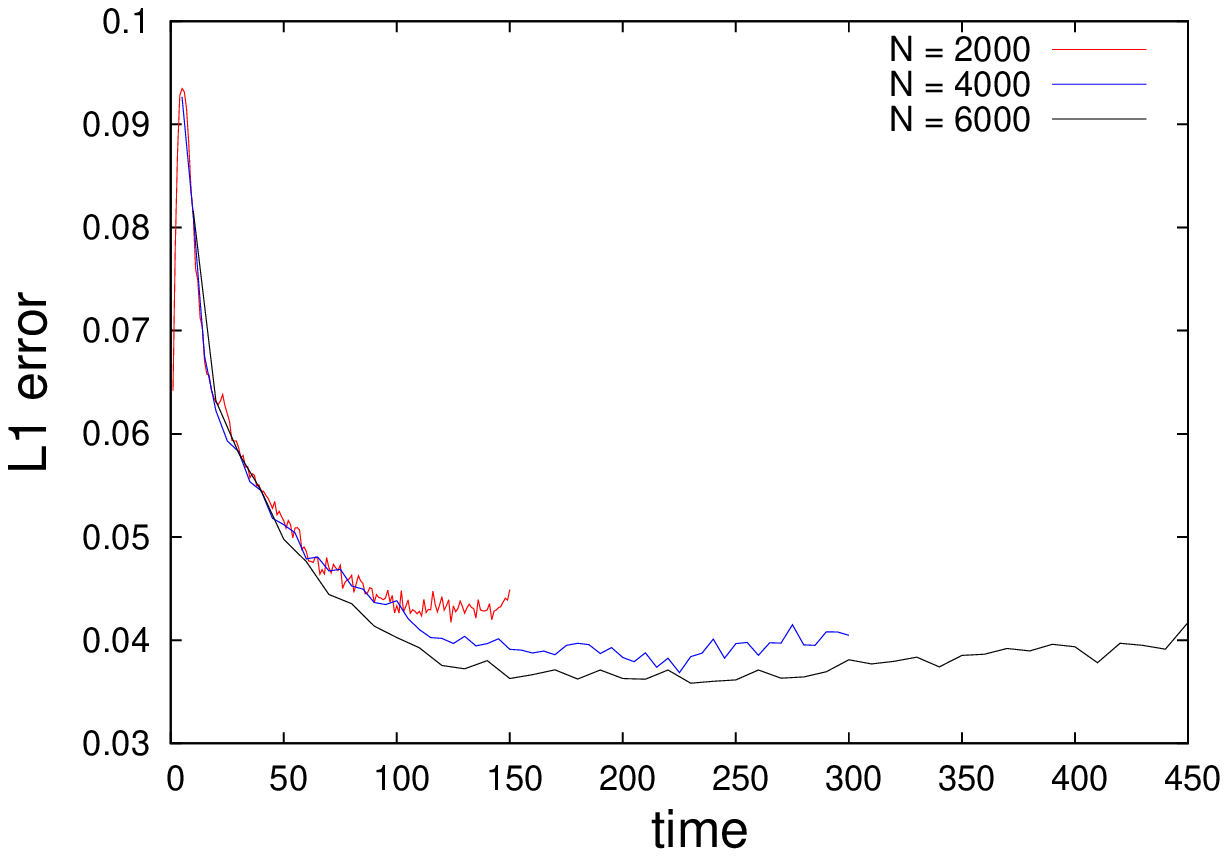}
\caption{\label{fig:cv_FPU_heat} (FPU potential, heat peak) Left: Optimal value of the scaling parameter for a given time, as given by~\eqref{eq:instantaneous_scaling_factor}. Right: $L^1$ error~\eqref{eq:minimization_pbm} for the optimal value of the parameters.}
\end{figure}

%-------- KVM ----------
\subsubsection{KVM potential}

The rescaled sound and heat peaks for the KvM potential are presented in Figure~\ref{fig:comparison_KVM}. The agreement with the predicted scaling function is again qualitatively excellent. On a quantitative level, the numerical parameters obtained by the minimization procedure are:
\begin{itemize}
\item for sound peaks, exponent $\delta_{\rm num} = 2/3$, fixed to its theoretical value since $\delta_{\rm num}$ turns out to be extremely close to 2/3, velocity $c_{\rm num} = -3.996$, compared to the theoretical value $c_{\rm theor} = -4$, and scaling factor $\lambda_{\rm s} \simeq 2.81$. The scaling factor is in excellent agreement with the theoretical value $\lambda_1 = 2\sqrt{2} |G^1_{11}| = 2.83$ predicted by~\eqref{eq:prediction_f1}. 
\item for heat peaks, we consider as a reference the maximally asymmetric Levy distribution with $\alpha = 1.57$ instead of $5/3$ since this value of the $\alpha$ parameter allows to further decrease the error~\eqref{eq:minimization_pbm}. We find a velocity $c_{\rm num} \simeq 0$, an exponent $\delta_{\rm num} = 0.633$, somewhat away from the theoretically predicted value $3/5$, and a scaling factor $\lambda_2 \simeq 2.51$. The latter value is quite off the theoretical value~$4.21$ predicted by~\eqref{eq:prediction_f2}.
\end{itemize}
% \lambda for heat : exec('compute_R_KvM.sce') et 'lambda_sound = 2*sqrt(2)*abs(Gu(2,2))' puis 'lambda_heat = (-c)^(-1/3)*(Ge(2,2)^2)*1.81/(lambda_sound)^(2/3)'
The evolution of the scaling factors as a function of the time index is reported in Figures~\ref{fig:cv_KVM_sound} and~\ref{fig:cv_KVM_heat}, together with the $L^1$ error. The behavior and orders of magnitude of the error are similar to what is observed with the FPU potential. We again see in this example that the convergence to the limiting regime for the sound peak is slightly faster than for the heat peak.

%obtained with RES/Sound/COMPARE_GNUPLOT_SOUND_KVM
% obtained with RES/Heat/COMPARE_GNUPLOT_HEAT_KVM
\begin{figure}[h]
\includegraphics[width=7.6cm]{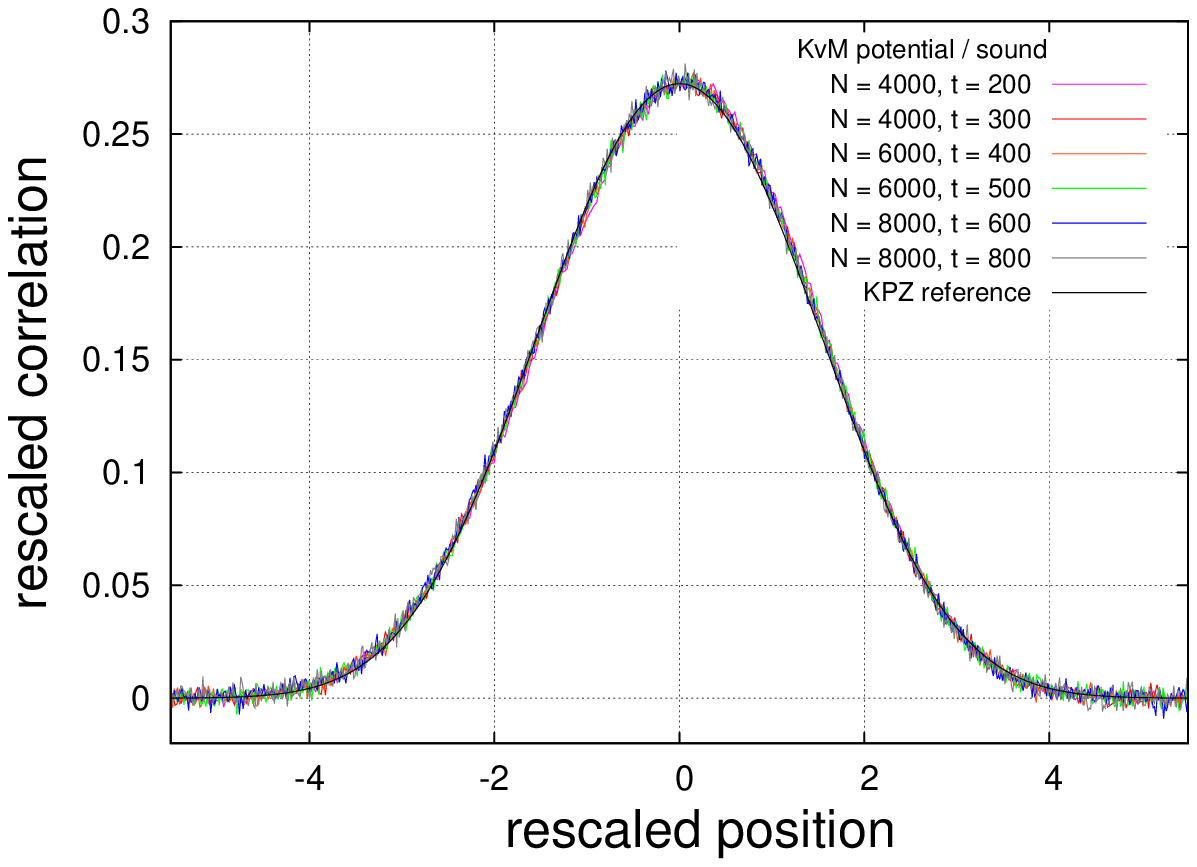}
\includegraphics[width=7.6cm]{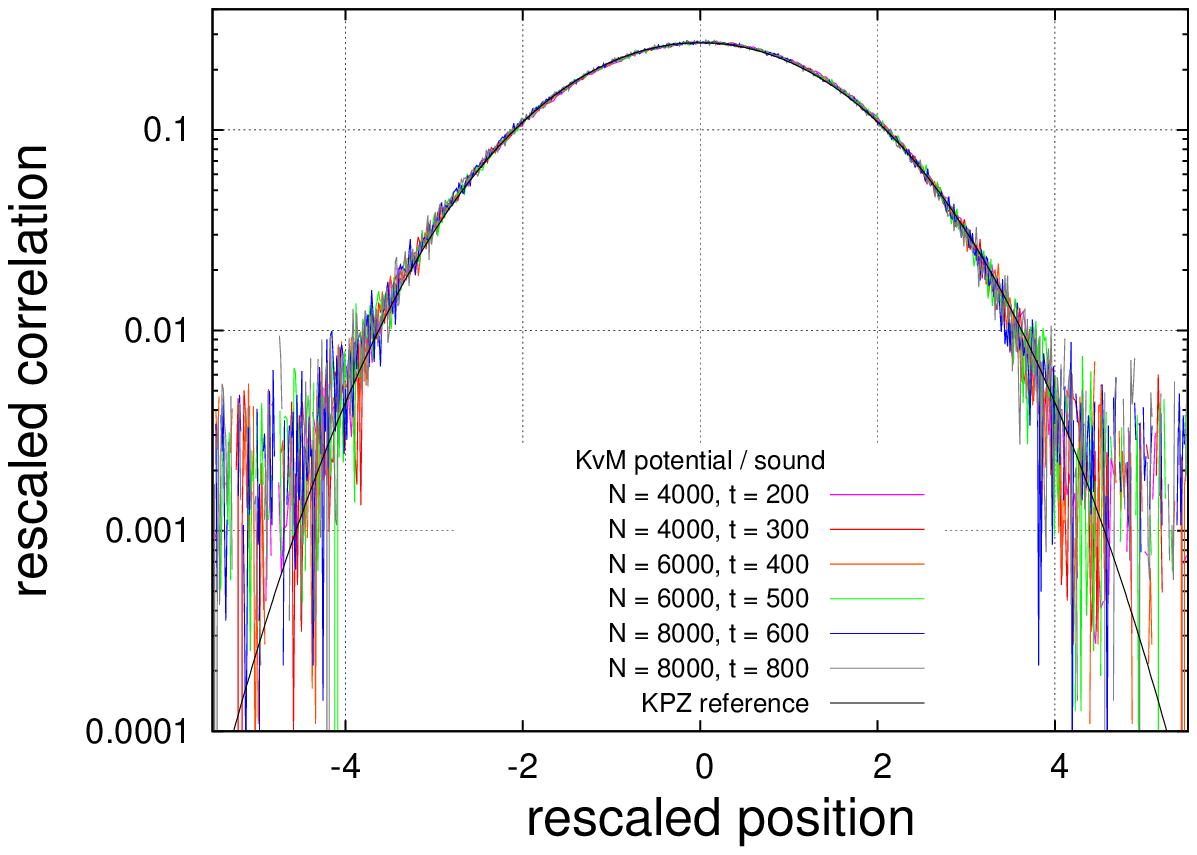}  \\
\includegraphics[width=7.6cm]{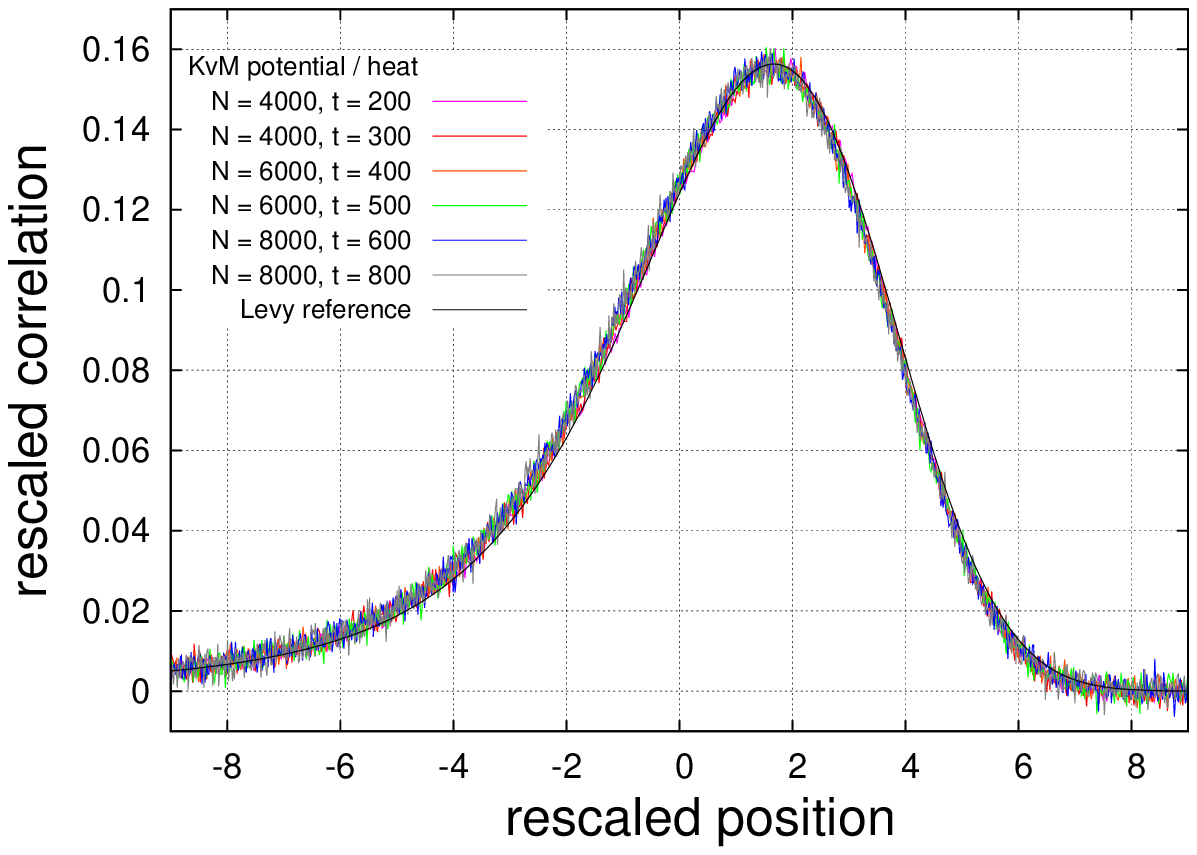}
\includegraphics[width=7.6cm]{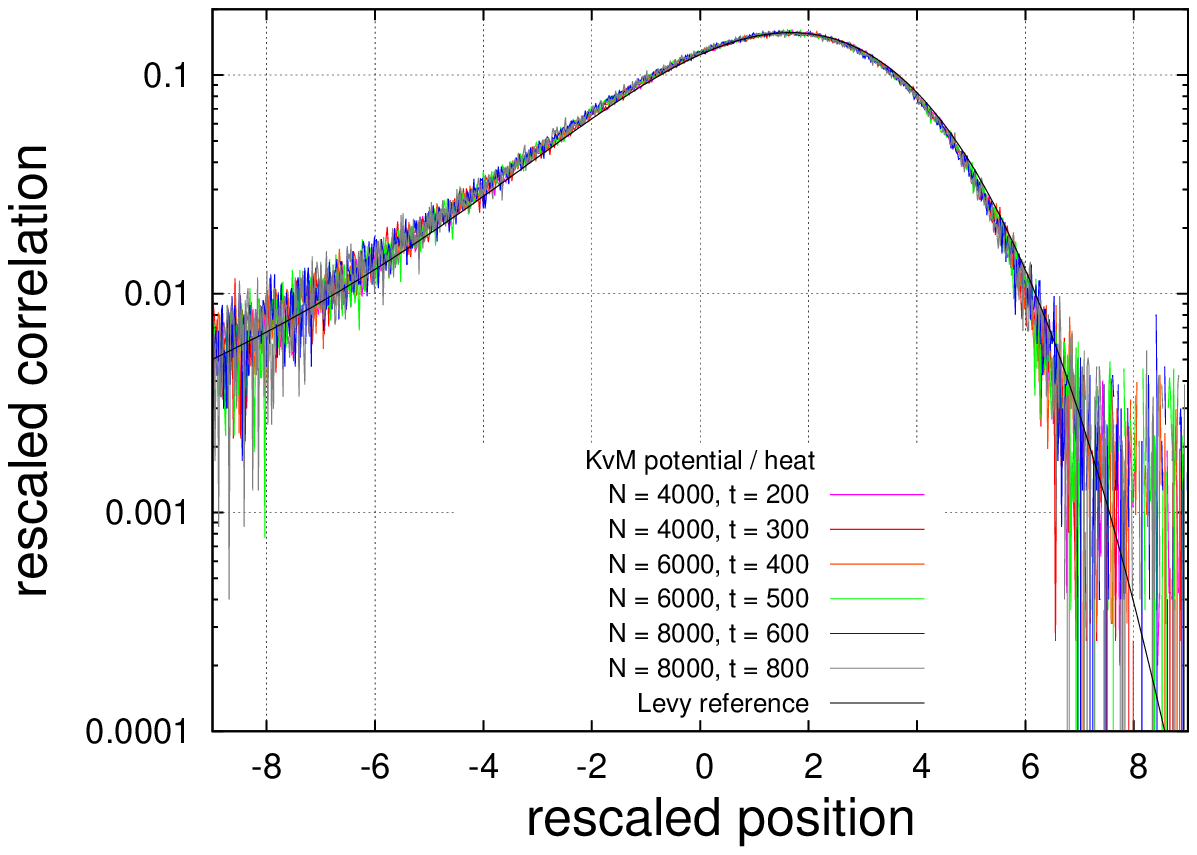}
\caption{\label{fig:comparison_KVM} (KvM potential)  Comparison of rescaled sound and heat peaks. The first line corresponds to sound modes, the second to heat modes. The reference for heat peaks is a maximally asymmetric Levy distribution with parameter $\alpha = 1.57$ instead of $1.67$. Logarithmic plots are provided in the right column.}
\end{figure}

\begin{figure}[h]
\includegraphics[width=7.5cm]{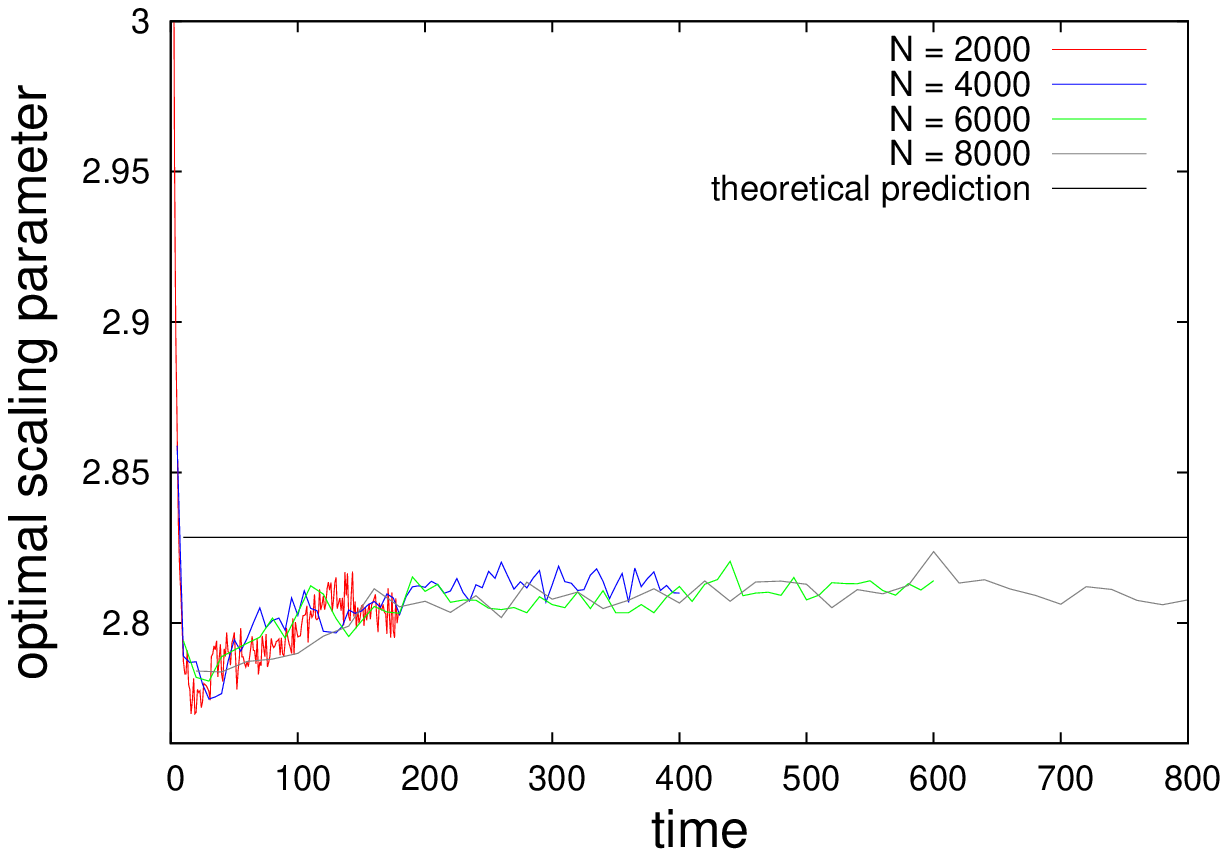}
\includegraphics[width=7.5cm]{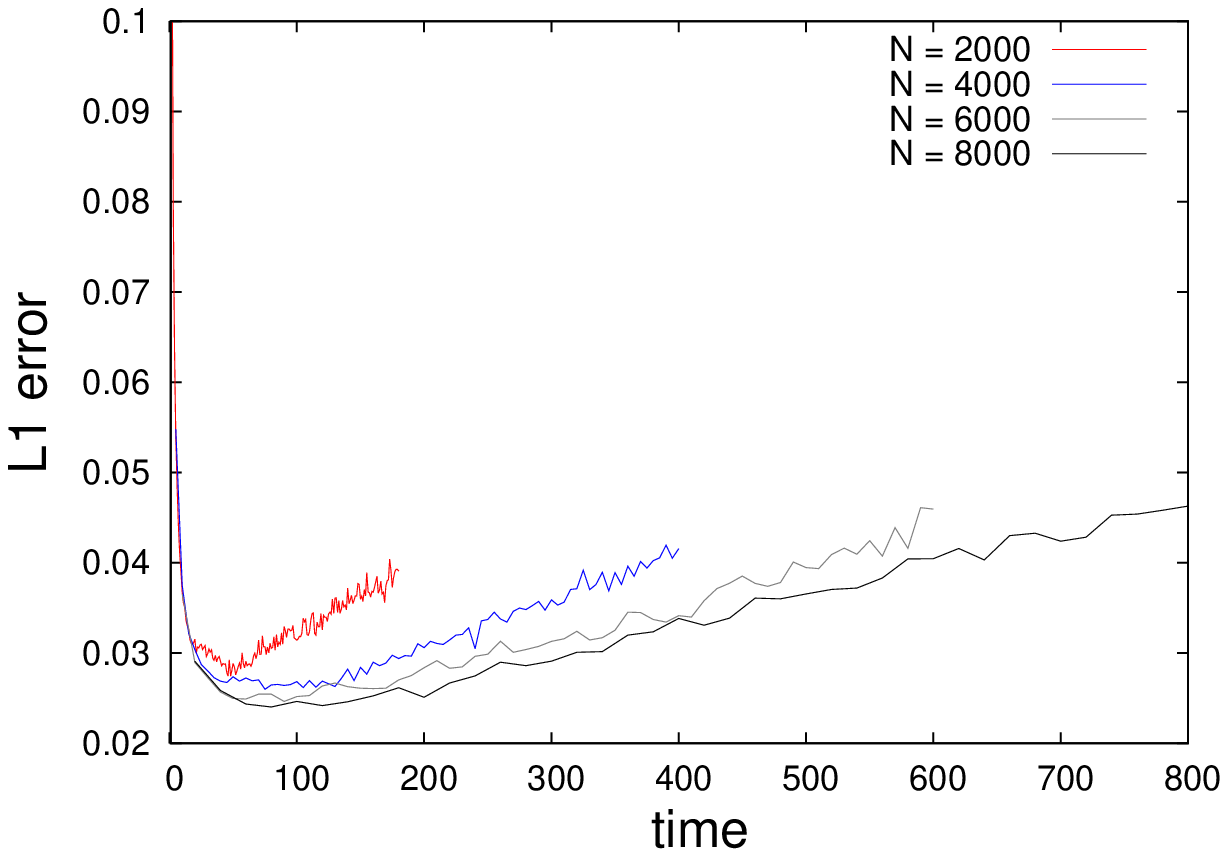}
\caption{\label{fig:cv_KVM_sound} (KvM potential, sound peak) Left: Optimal value of the scaling parameter for a given time, as given by~\eqref{eq:instantaneous_scaling_factor}. Right: $L^1$ error~\eqref{eq:minimization_pbm} for the optimal value of the parameters.}
\end{figure}

\begin{figure}[h]
\includegraphics[width=7.5cm]{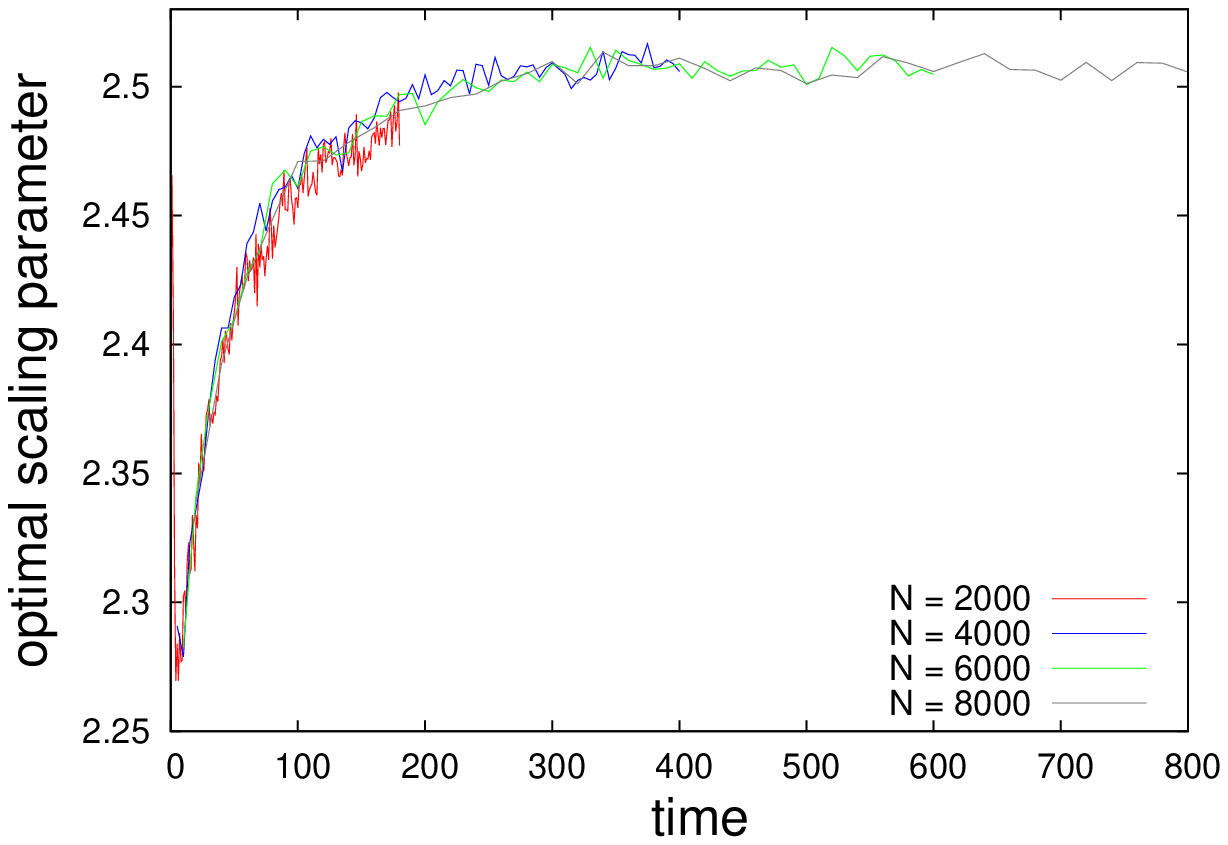}
\includegraphics[width=7.5cm]{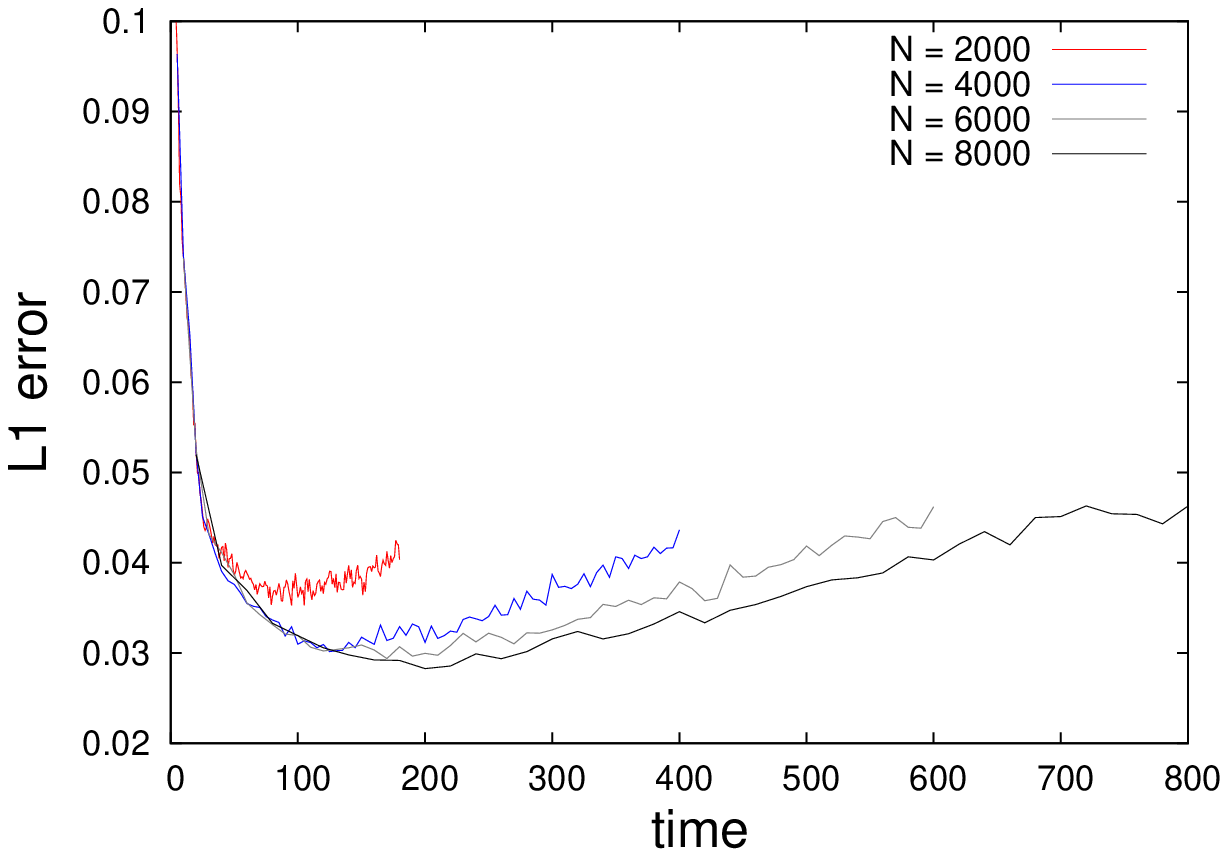}
\caption{\label{fig:cv_KVM_heat} (KvM potential, heat peak) Left: Optimal value of the scaling parameter for a given time, as defined by~\eqref{eq:instantaneous_scaling_factor}. Right: $L^1$ error~\eqref{eq:minimization_pbm} for the optimal value of the parameters.}
\end{figure}

%---------------- conclusion -----------------------
\section{Conclusions}\label{sec5}
\setcounter{equation}{0}
Comparable simulations have been carried out for a two-component stochastic lattice gas \cite{FeSS13} and for Hamiltonian anharmonic chains with hard core collisions \cite{MS14}, resp. with an asymmetric FPU potential \cite{DDSMS14,St14}. The latter two models have three conserved fields. The lattice gas has two KPZ peaks with distinct velocities. The agreement with KPZ is of a similar quality as obtained here, including the values for the non-universal coefficients. It could be that the $G$-matrices
were such as to favor small finite time corrections. On the other hand, for the anharmonic chains the agreement is less perfect. For hard-core collisions the predicted shape of the peaks is achieved with an $L^1$ error of the order of $3\%$, but the non-universal coefficients deviate 
considerably from their predicted values. Such deviations are even more pronounced for the FPU chains. For instance, at the largest time and size available, the sound peaks still show a slight asymmetry.

It is remarkable that the space-time correlation functions obtained from numerical simulations of the BS model are in such a good agreement with the ones of nonlinear fluctuating hydrodynamics. 

KPZ scaling is based on decoupling and is expected to be exact for sufficiently long times.  On the other hand, the Levy distribution is based on mode-coupling, which is an approximation. As also observed in other models, for our simulations  the fit to the $\tfrac{5}{3}$-Levy distribution is so precise that one is tempted to conjecture it  to be the true long time scaling function.

%----------------- APPENDIX ---------------------
\begin{appendix}
\section{Scaling functions for two cross-coupled modes}
\label{secA}
\setcounter{equation}{0}

We study the asymptotic behavior  of two cross-coupled Burgers equation of the form 
\begin{equation}\label{1}
\partial_t u_{\sigma}+\partial_x\big(\sigma c u_{\sigma} + \lambda(u_{-\sigma})^2 -D\partial_x u_{\sigma} +\sqrt{2D}\xi_{\sigma}\big)=0\,,\quad \sigma = \pm 1,
\end{equation}
for velocity $c >0$, diffusion constant $D>0$, and strength of nonlinearity $\lambda > 0$. This is the case ``\textit{gold}-Levy'' from Table 3, row 1, in Section \ref{sec2} (with the simplification that the strength of the nonlinearity is assumed to be the same for both modes). Note that, compared to~\eqref{2.3} the index of the modes is $\pm 1$ instead of $1,2$ and the frame of reference is such that the modes have opposite velocities.

In the diagonal approximation, compare with (\ref{2.5}) - (\ref{2.6a}), the respective mode-coupling equations read 
\begin{equation}
  \label{2}
  \partial_t f_\sigma(x,t) = \left(-\sigma c \partial_x+D \partial^2_x\right) f_\sigma(x,t)+ 2\lambda^2 \int^t_0 \int_{\mathbb{R}} f_\sigma(x-y,t-s) \partial^2_y\big( f_{-\sigma}(y,s)^2\big)dy \, ds .
\end{equation}
Initially $f_\sigma(x,0) = \delta(x)$ and the normalization is preserved,
\begin{equation}
\label{eq:normalization_f}
\int_\mathbb{R} f_{\sigma}(x,t) \, dx = 1.
\end{equation}
Furthermore, by symmetry of the equations, 
\begin{equation}
  \label{eq:symmetry_f_sigma}
  f_{\sigma}(x,t)= f_{-\sigma}(-x,t).
\end{equation}

Our goal is to find the self-similar solution to~\eqref{2}. We will establish that the appropriate space-time scaling
is $x/t^{1/\gamma}$ with $\gamma$ the golden mean,
\begin{equation}
  \label{eq:golden_mean}
  \gamma = \frac{1 + \sqrt{5}}{2} \simeq 1.618.
\end{equation}
 The scaling function turns out to be  the maximally asymmetric $\gamma$-Levy distribution, see~\eqref{eq:final_form_scaling_function} below and Section~\ref{sec:properties_scaling_fcts} for a discussion of its tail properties.

\subsection{Equation for the scaling functions}
\label{sec:derivation_eq_scaling}

We use the same Fourier transform conventions as in~\cite{Spohn14},
\[
\hat{g}(k) = \int_\mathbb{R} g(x) \, \rme^{-2\ri\pi kx} \, dx.
\]
Taking the spatial Fourier transform of~\eqref{2} leads to
\begin{equation}
\label{eq:Fourier_mc}
\begin{aligned}
\partial_t \hat{f}_\sigma(k,t) & = -\left(2\ri\pi\sigma c k + (2\pi k)^2 D \right) \hat{f}_\sigma(k,t) \\ 
& \quad - 2(2\pi k)^2 \lambda^2 \int^t_0 \hat{f}_\sigma(k,t-s) \Big( \int_{\mathbb{R}} \hat{f}_{-\sigma}(k-q,s)\, \hat{f}_{-\sigma}(q,s) \, dq \Big) ds .
\end{aligned}
\end{equation}
We assume that, relative to $\sigma ct$, $f_\sigma$ is  a self-similar solution with still to be determined space-time scale. Recall that if a function~$f$ is self-similar,
\[
f(x,t) = t^{-a}F(t^{-a}(x\mp ct)),
\]
then $\hat{f}(k,t) = \mathrm{e}^{\mp 2\ri\pi kct} \hat{F}(k t^a)$. We therefore make the following scaling ansatz
\begin{equation}\label{eq:ansatz_self_similar}
\hat{f}_1(k,t) = \mathrm{e}^{-2\ri\pi kct} h(k^\gamma t)\,, \qquad \hat{f}_{-1}(k,t) = \mathrm{e}^{2\ri\pi kct} g(k t^\beta),
\end{equation}
which is expected to be valid asymptotically only, as made precise in~\eqref{eq:limiting_meaning}.

We consider the forcing exerted by $f_{-1}$ on~$f_{1}$, which amounts to regarding the function~$g$ as the input and $h$ as the output. Let us first state some properties of the functions $g,h$. Since  $f_\sigma$ is real-valued, $h(-w) = \overline{h(w)}$ and $g(-w) = \overline{g(w)}$. We therefore restrict ourselves to $k >0$ in the sequel.
% cf. f_+(x,t) = F((x-ct)/t^a) and do the change of variables x <- x+ct in the Fourier transform
%Note also that, by symmetry, $\hat{f}_1(k,t) = \overline{\hat{f}_{-1}(k,t)}$, so that 
%\begin{equation}\label{eq:symmetry_h_g}
%  h(k^\gamma t) = \overline{ g\left(\left(k^{1/\beta}t\right)^\beta\right) }\,,
%\end{equation}
Plugging the ansatz~\eqref{eq:ansatz_self_similar} into~\eqref{eq:Fourier_mc}, the equation for $\hat{f}_1$ turns into 
\[
k^\gamma h'(k^\gamma t) = -(2\pi k)^2 D h(k^\gamma t) - 2(2\pi k)^2 \lambda^2 \int^t_0 h\big(k^\gamma(t-s)\big) \int_\mathbb{R} g\big((k-q)s^\beta\big) g\big(q s^\beta\big)\rme^{4\ri\pi cks} \, dq \, ds ,
\]
so that, introducing the new variables $w = k^\gamma t$ and $u = q s^\beta$,
\[
\begin{aligned}
h'(w) & = -4\pi^2D\, k^{2-\gamma} h(w) \\
& \ \ - 8\pi^2 \lambda^2 \, k^{2-\gamma} \int^{k^{-\gamma} w}_0 \rme^{4\ri\pi cks}s^{-\beta} \, h\left(w-k^\gamma s\right) \Big(\int_\mathbb{R} g\left(ks^\beta-u\right) g\left(u\right) \, du\Big) ds .
\end{aligned}
\]
Here $t$ has been eliminated and we study the limit $k \to 0$. We rescale the time integration variable as $s = k^{-a} \theta$ and obtain
\[
\begin{aligned}
& h'(w) = -4\pi^2D\, k^{2-\gamma} h(w) \\
& - 8\pi^2 \lambda^2 \, k^{2-\gamma+a(\beta-1)} \int^{k^{a-\gamma}w}_0 \rme^{4\ri\pi ck^{1-a}\theta}\theta^{-\beta} \, h\left(w-k^{\gamma-a} \theta\right) \Big(\int_\mathbb{R} g\left(k^{1-a\beta}\theta^\beta-u\right) g\left(u\right) \, du\Big) d\theta  .
\end{aligned}
\]
The choice $a=1$ is the only one leading to a non-trivial limit in the integral over~$\theta$ as $k\to 0$. Indeed, for $a < 1$, the exponential factor converges to~1 and the integrand is proportional to $\theta^{-\beta}$ which is not integrable over 
$\mathbb{R}_+$, while for $a > 1$ the integral converges to~0, since the exponential factor oscillates more and more. Setting 
\begin{equation}
\label{eq:choice_gamma_1}
\gamma = 1 + \beta, \qquad 0< \beta < 1,
\end{equation}
one arrives at
\[
\begin{aligned}
h'(w) & = -4\pi^2 D\, k^{2-\gamma} h(w) \\
& \quad - 8\pi^2 \lambda^2 \, \int^{k^{1-\gamma}w}_0 \rme^{4\ri\pi c \theta}\theta^{-\beta} \, h\left(w-k^{\beta} \theta\right) \Big(\int_\mathbb{R} g\left(k^{1-\beta}\theta^\beta-u\right) g\left(u\right) \, du\Big) d\theta .
\end{aligned}
\]
In the limit $k\to 0$,
\begin{equation}\label{4}
  h'(w) = - h(w) (4\pi\lambda)^2 \Big(\int_0^\infty |g(u)|^2 du\Big) \Big(\int_0^\infty \mathrm{e}^{4\mathrm{i}\pi c \theta}\theta^{-\beta}\,d\theta\Big),
\end{equation}
which determines $h$ once the values of the integrals on the right-hand side are known. We can now make precise the meaning of the limiting procedure, namely 
\begin{equation}
  \label{eq:limiting_meaning}
  \lim_{k \to 0} \mathrm{e}^{2\ri\pi c k^{1-\gamma} w } \hat{f}_1\left(k,k^{-\gamma}w\right) = h(w),
\end{equation}
where $h$ is the solution of~\eqref{4}.

\subsection{Cross-coupled scaling functions}

The time integral in~\eqref{4} can be computed analytically as (see~\cite[Section~6.33]{zwillinger})
% and using \Gamma(x) \Gamma(1-x) = \pi/\sin(\pi x) and various other trigonometric relations to reformulate the ratio \cos(\pi(1-\beta)/2)/\sin(\pi \beta) as 1/(2\cos(\pi\beta/2))
\begin{equation}
\label{eq:value_time_integral}
\int_0^\infty \mathrm{e}^{4\mathrm{i}\pi c \theta}\theta^{-\beta}\,d\theta = (4\pi c)^{-1+\beta} \int_0^\infty \mathrm{e}^{\mathrm{i} s}s^{-\beta}\,ds = a \Big(1 + \frac{\ri}{\tan(\pi\beta/2)}\Big)
\end{equation}
with
\[
a = (4\pi c)^{-1+\beta} \, \frac{\pi}{2\Gamma(\beta) \cos(\pi\beta/2)}.
\]
We repeat now the derivation in Section~\ref{sec:derivation_eq_scaling} considering~$h$ as input and~$g$ as the output. By symmetry~\eqref{eq:symmetry_f_sigma}, one concludes that
\[
h(k^\gamma t) = \overline{g\big((k^{1/\beta} t)^\beta\big)}, 
\]
which forces $h(w) = \overline{g(w^\beta)}$ and 
\[
\gamma = \frac1\beta.
\]
Combined with~\eqref{eq:choice_gamma_1}, this implies that $\gamma$ equals the golden mean~\eqref{eq:golden_mean}. The normalization condition~\eqref{eq:normalization_f} implies $h(0) = 1$. Hence, noting that $\tan(\pi\beta/2) = -1/\tan(\pi\gamma/2)$,
\begin{equation}
\label{eq:expression_h_first_round}
h(w) = \exp\left(-(4\pi\lambda)^2 a\big(1-\mathrm{i}\tan(\pi\gamma/2)\big) A w\right),
\end{equation}
with 
\[
A =  \int_0^\infty |g(u)|^2 du =  \int _0^\infty |h(w^\gamma)|^2 dw.
\]
Inserting~\eqref{eq:expression_h_first_round} in the latter expression, it follows that
\begin{equation}\label{9}
A = (a (4\pi \lambda)^2)^{-1/\gamma^2}\widetilde{A}^{1/\gamma},
\qquad \widetilde{A}
= \int_0^\infty \mathrm{e}^{-2 w ^{\gamma}}\,dw = \frac{2^{-1/\gamma}}{\gamma} \Gamma\left(\frac1\gamma\right).
\end{equation}
Thereby we finally obtain the scaling function
\begin{equation}
  \label{eq:final_form_scaling_function}
f_\sigma(k,t) = \exp\left(-2\mathrm{i}\pi \sigma k c t -  C\,|2\pi k|^\gamma
\big( 1 -\mathrm{i}\sigma \mathrm{sgn}(k)\tan(\pi\gamma/2) \big) t \right),
\end{equation}
with
% C = (a\widetilde{A}(4\pi\lambda)^2)^{1/\gamma}
\[
C = \frac12 \lambda^{2/\gamma} \left(\frac{1}{\gamma \sin(\pi\gamma/2)}\right)^{1/\gamma} c^{1-2/\gamma},
\]
which one recognizes as the Fourier transform of an $\alpha$-stable law with $\alpha = \gamma$ and maximal asymmetry $b = \sigma$, see Section \ref{sec:properties_scaling_fcts} below. This expression reduces to the more general expression derived in~\cite{PSS14a} in the case when the strengths of the nonlinearities for the cross-coupled modes are different.

For two components the Levy distribution is necessarily maximal asymmetric. For three modes,
there is the possibility to sandwich the Levy peak inbetween two sound peaks with a rapid fall off, as KPZ or Gaussian. 
Then the Levy distribution could be partially asymmetric with the tails cut off at the location of the sound peaks.
Such a situation is realized in all anharmonic chains. Since the two sound peaks are mirror images relative to $0$,
the Levy distribution turns out to be symmetric, $b = 0$. On a mathematical level, the only result available is the harmonic chain with random velocity exchanges. In this case the sound peaks are Gaussian and the heat peak is
Levy with parameters $\alpha = \tfrac{3}{2}$ and $b = 0$ \cite{JKO14}. 

%------------------------------------------------------------------------------------
\subsection{Levy distributions and their asymptotic properties}
\label{sec:properties_scaling_fcts}
The Levy distributions are defined through their Fourier transform as
\begin{equation}
\label{eq:generic_Levy}
f_{{\rm Levy},\alpha,b}(x) = \frac{1}{2\pi} \int_\mathbb{R} \varphi_{\alpha,b}(k) \, \mathrm{e}^{\mathrm{i} kx} \, dk, 
\quad
\varphi_{\alpha,b}(k) = \exp\big(-|k|^\alpha \big[1-\mathrm{i} b \tan(\tfrac{1}{2}\pi \alpha) \mathrm{sgn}(k)\big]\big).
\end{equation}
There are two parameters: $\alpha$  controls the steepness, $0 < \alpha < 2$, and $b$  controls the asymmetry, 
$|b| \leq 1$. At the singular point $\alpha = 2$, $b = 0$ the distribution is Gaussian. For $|b| > 1$ the Fourier integral no longer defines a non-negative function. If $|b|< 1$, the asymptotic decay of  $f_{{\rm Levy},\alpha,b}(x)$  is determined by $\alpha$ and is given by $|x|^{-\alpha-1}$ for $|x| \to \infty$. At $|b| = 1$ the two tails show different decay. The functions corresponding to $b = 1$ and $b = -1$ are mirror images, for $b=1$ the slow decay being for $x \to -\infty$ and still as $|x|^{-\alpha-1}$. For $0 < \alpha \leq 1$, $f_{{\rm Levy},\alpha,1}(x) = 0$ for $x > 0$, while for $1 < \alpha < 2$ the decay becomes stretched exponential as
$\exp(-c_0 x^{\alpha/(1 - \alpha)})$ with known constant $c_0$. We refer to~\cite{UZ99} for more details.
In our context only the maximal asymmetry $b = \pm 1$ with $1 <\alpha < 2$
is realized. 

%------------------- mod KPZ --------------------
\section{Modified KPZ scaling}
\label{sec:modified_KPZ}

In this section we study modified KPZ from Table 2, row 2 of Section \ref{sec2}, in which case $G^1_{11} \neq 0$ and mode~2 is diffusive, but has a non-trivial feedback to mode~1 since $G^1_{22} \neq 0$. More precisely, upon changing the frame of reference, we assume that
\[
f_2(x,t) = \frac{1}{\sqrt{4\pi D t}} \, \rme^{-(x + ct)^2/4D t},
\]
$c > 0$, while $f_1$ evolves according to
\[
\begin{aligned}
\partial_t f_1 = D_1 \partial_x^2f_1 & + 2 \left(G^1_{11} \right)^2 \int^t_0 \int_{\mathbb{R}} 
\partial^2_xf_1(x-y,t-s) f_1(y,s)^2 \, dy \, ds  \\
& + 2 \left(G^1_{22} \right)^2 \int^t_0 \int_{\mathbb{R}}  \partial^2_xf_1(x-y,t-s) f_2 (y,s)^2 \, dy \, ds,
\end{aligned}
\]
compare with~\eqref{2.5}-\eqref{2.6a}. Through Fourier transform in space one obtains
\[
\begin{aligned}
\partial_t \hat{f}_1(k,t) & =  - D_1 (2\pi k)^2   \hat{f}_1(k,t) \\
& \quad - 2(2\pi k)^2 \left(G^1_{11} \right)^2 \int^t_0 \hat{f}_1(k,t-s) \Big( \int_{\mathbb{R}} \hat{f}_1(k-q,s)\, \hat{f}_1(q,s) \, dq \Big) ds \, \\
& \quad - 2(2\pi k)^2 \left(G^1_{22} \right)^2 \int^t_0 \hat{f}_1(k,t-s) \Big( \int_{\mathbb{R}} \hat{f}_2(k-q,s)\, \hat{f}_2(q,s) \, dq \Big) ds .
\end{aligned}
\]
As in Section~\ref{sec:derivation_eq_scaling}, it suffices to consider $k>0$. Following the scheme in~\cite[Section~4]{Spohn14}, we make the ansatz
\begin{equation}
  \label{eq:ansatz_mod_KPZ1}
  f_1(k,t) =  F\big((\lambda_\mathrm{s} t)^{2/3} k\big)
\end{equation}
with $\lambda_\mathrm{s} = 2 \sqrt{2} |G^1_{11}|$. Setting momentarily $G^1_{22} = 0$, and substituting $ u = (\lambda_\mathrm{s} t)^{2/3} k$, one arrives at 
\begin{equation}
\label{B1}
\tfrac{2}{3}u F'(u) = - \pi^2 u^2 \int^1_0 F\big((1-\theta)^{2/3} u\big) \Big( \int_{\mathbb{R}} F\big(\theta^{2/3} (u-v)\big)\,F\big(\theta^{2/3} v\big)  \, dv \Big) d\theta.
\end{equation}

Next we set momentarily $G^1_{11} = 0$. Then we are back to the problem discussed in Section \ref{sec:derivation_eq_scaling} with $\beta =1/2$, $\gamma = 3/2$ and input function  $\hat{f}_{2}(k,t) = \mathrm{e}^{2\ri\pi kct} g(k t^{1/2})$ with $g(k) = \exp\big(-D(2\pi k)^2\big)$. In the scaling limit the output function is $h(k^{3/2}t)$, which satisfies 
\[
h'(w) = - h(w) (4\pi G^1_{22})^2 \Big(\int_0^\infty |g(u)|^2 du\Big) \Big(\int_0^\infty \mathrm{e}^{2\mathrm{i}\pi c \theta}\theta^{-1/2}\,d\theta\Big).
\]
Working out the integrals yields
\[
h'(w) = - h(w) (4\pi G^1_{22})^2 ( 4\sqrt{\pi D})^{-1} (1 + \mathrm{i})(2 \sqrt{c})^{-1}.
\]
Since $w = k^{3/2}t$, one concludes $h(w) = F\big((\lambda_\mathrm{s}w)^{2/3}\big)$. The linear equation  $h'(w) = ah(w)$ translates into
\begin{equation}
\label{B2}
\tfrac{2}{3} F'(u) =  a (\lambda_\mathrm{s})^{-1} \sqrt{u} F(u).
\end{equation}

Combining~(\ref{B1}) and (\ref{B2})  one arrives at the fixed point equation for the scaling function $F$,
\begin{equation}
\label{modKPZ}
\begin{aligned}
\tfrac{2}{3}  \, F'(u) & = - \pi^2 u \int^1_0 F\big((1-\theta)^{2/3} u\big) \left( \int_{\mathbb{R}} F\big(\theta^{2/3} (u-v)\big)\,F\big(\theta^{2/3} v\big)  \, dv \right) d\theta \, \\
& \ \ \ -  (4\pi G^1_{22})^2 ( 4 \sqrt{\pi D})^{-1} (1 + \mathrm{i})(2 \sqrt{c})^{-1} (2 \sqrt{2}|G_{11}^1|)^{-1} \, \sqrt{u} F(u)  .
\end{aligned}
\end{equation}
If $G^1_{22} =0$, then \eqref{modKPZ} reduces to the fixed point equation for $f_{\rm KPZ}$ in the mode-coupling 
approximation. Now a term linear in $F$ is added. Presumably this results in a one-parameter family of scaling functions,
depending on the prefactor of the linear term. Most likely such  a behavior persists for the true coupled Burgers equations. 

%----------------------------- coupling constants ------------------
\section{Expressions for the coupling constants}
\label{sec:coupling_constants}

We follow here the strategy presented in~\cite[Appendix~A]{Spohn14} to compute the various coefficients appearing in  the mode-coupling equations. Mode~1 corresponds to the sound mode, while mode~2 represents the heat mode.

For three random variables $\mathcal{A},\mathcal{B},\mathcal{C}$, we denote the third cumulant by
\[
\begin{aligned}
& \left \langle \mathcal{A};\mathcal{B};\mathcal{C}\right\rangle_{\tau,\beta} = \left \langle \mathcal{A}(\eta_0) \mathcal{B}(\eta_0) \mathcal{C}(\eta_0) \right\rangle_{\tau,\beta} \\
& \qquad  - \left \langle \mathcal{A}(\eta_0)\mathcal{B}(\eta_0)\right\rangle_{\tau,\beta}\left \langle \mathcal{C}(\eta_0)\right\rangle_{\tau,\beta} - \left \langle \mathcal{A}(\eta_0)\mathcal{C}(\eta_0)\right\rangle_{\tau,\beta}\left \langle \mathcal{B}(\eta_0)\right\rangle_{\tau,\beta} - \left \langle \mathcal{B}(\eta_0)\mathcal{C}(\eta_0)\right\rangle_{\tau,\beta}\left \langle \mathcal{A}(\eta_0)\right\rangle_{\tau,\beta} \\
& \qquad +2 \left \langle \mathcal{A}(\eta_0) \right\rangle_{\tau,\beta}\left \langle \mathcal{B}(\eta_0) \right\rangle_{\tau,\beta}\left \langle \mathcal{C}(\eta_0) \right\rangle_{\tau,\beta}.
\end{aligned} 
\]

\subsection{Matrix $R$ and sound velocity}

The right eigenvectors of the matrix $A$ are proportional to
\[
\psi_1 = Z_1^{-1} \begin{pmatrix} 1 \\ -\tau \end{pmatrix}, 
\qquad
\psi_2 = Z_2^{-1} \begin{pmatrix} \partial_e \tau \\ -\partial_h \tau \end{pmatrix}, 
\]
with, respectively, associated eigenvalues 0 and
\[
c = 2 (\partial_h  -\tau \partial_e) \tau.
\]
 The corresponding left eigenvectors are proportional to
\[
\widetilde{\psi}_1 = \widetilde{Z}_1^{-1} \begin{pmatrix} \partial_h \tau \\ \partial_e \tau \end{pmatrix}.
\qquad
\widetilde{\psi}_2 = \widetilde{Z}_2^{-1} \begin{pmatrix} \tau \\ 1 \end{pmatrix}, 
\]
The coefficients $\widetilde{Z}_1,\widetilde{Z}_2$ are obtained from the diagonal conditions $RCR^\mathrm{T} = 1$, the $R$ matrix being constructed from the left eigenvectors. The coefficients $Z_1,Z_2$ are determined by the condition $R R^{-1} = 1$, with the inverse $R^{-1}$ constructed from the right eigenvectors. By some computations one obtains
\begin{equation}
\label{eq:def_c}
c = -2 \Gamma^{-1}\langle V + \tau \eta; V + \tau \eta\rangle_{\tau,\beta} <0, 
\qquad
\Gamma = \beta \big( \left\langle \eta ; \eta\right\rangle_{\tau,\beta}\left\langle V ; V\right\rangle_{\tau,\beta} - \left\langle \eta; V\right\rangle_{\tau,\beta}^2 \big),
\end{equation}
as well as
\[
\widetilde{Z}_1 = \sqrt{-\frac{c}{2\beta}},
\qquad
\widetilde{Z}_2 = \sqrt{-\frac{\Gamma c}{2}}.
\]
Moreover,
\begin{equation}
\label{eq:matrix_R}
R = \begin{pmatrix} 
\partial_h \tau/\widetilde{Z}_1 & \partial_e \tau/\widetilde{Z}_1 \\ 
\tau/\widetilde{Z}_2 & 1/\widetilde{Z}_2 \\
\end{pmatrix},
\end{equation}
with
\begin{equation}
\label{eq:partial_derivatives_tau}
\partial_h \tau = -\Gamma^{-1}\left\langle V ; V + \tau \eta\right\rangle_{\tau,\beta}, 
\qquad
\partial_e \tau = \Gamma^{-1}\left\langle \eta ; V + \tau \eta\right\rangle_{\tau,\beta}.
\end{equation}
Finally,
\[
Z_1 = \frac{c}{2\widetilde{Z}_1} = -\sqrt{-\frac{\beta c}{2}},
\qquad
Z_2 = -\frac{c}{2\widetilde{Z}_2} = \sqrt{-\frac{c}{2\Gamma}}. 
\]

\subsection{Hessians and coupling matrices $G$}
\label{sec:G_matrices}

The Hessians of the currents $j_h = 2\tau$ and $j_e = -\tau^2$ are 
\[
H_h = \begin{pmatrix} 
  \partial_h^2 j_h & \partial_h \partial_e j_h \\
  \partial_h \partial_e j_h & \partial_e^2 j_h \\
\end{pmatrix} = 2\begin{pmatrix} 
  \partial_h^2 \tau & \partial_h \partial_e \tau \\
  \partial_h \partial_e \tau & \partial_e^2 \tau \\
\end{pmatrix} , 
\]
and
\[
H_e = \begin{pmatrix} 
  \partial_h^2 j_e & \partial_h \partial_e j_e \\
  \partial_h \partial_e j_e & \partial_e^2 j_e \\
\end{pmatrix} = -\tau H_h - 2 \widehat{H}_e, 
\qquad 
\widehat{H}_e = \begin{pmatrix} 
(\partial_h \tau)^2 & \partial_h \tau  \partial_e \tau \\
\partial_h \tau  \partial_e \tau & (\partial_e \tau)^2 \\
\end{pmatrix}.
\]
The second derivatives of~$\tau$ with respect to $h,e$, which are required in order to evaluate the Hessian matrices $H_h,H_e$, are obtained by inverting the following systems,
\[
\begin{pmatrix} \partial_\tau h & \partial_\tau e \\ \partial_\beta h & \partial_\beta e \end{pmatrix} \begin{pmatrix} \partial_h^2 \tau \\ \partial_h \partial_e \tau \end{pmatrix} = \begin{pmatrix} \partial_\tau (\partial_h \tau) \\ \partial_\beta (\partial_h \tau) \end{pmatrix},  
\qquad
\begin{pmatrix} \partial_\tau h & \partial_\tau e \\ \partial_\beta h & \partial_\beta e \end{pmatrix} \begin{pmatrix} \partial_h\partial_e \tau \\ \partial^2_e \tau \end{pmatrix} = \begin{pmatrix} \partial_\tau (\partial_e \tau) \\ \partial_\beta (\partial_e \tau) \end{pmatrix},
\]
and using the expressions~\eqref{eq:partial_derivatives_tau} for the partial derivatives $\partial_h \tau,\partial_e \tau$, as well as the rules 
\[
\partial_\tau \left \langle \mathcal{A};\mathcal{B}\right\rangle_{\tau,\beta} = -\beta \left \langle \mathcal{A};\mathcal{B};\eta \right\rangle_{\tau,\beta}, 
\qquad
\partial_\beta \left \langle \mathcal{A};\mathcal{B}\right\rangle_{\tau,\beta} = -\left \langle \mathcal{A};\mathcal{B};V+\tau\eta \right\rangle_{\tau,\beta}.
\]
The elements of the $G$ matrices are then computed as
\[
\begin{aligned}
G_{\alpha\alpha'}^1 & = \tfrac{1}{2} \Big( R_{11}\big( \psi_\alpha^\mathrm{T} \cdot H_h \psi_{\alpha'}\big) + R_{12}\big( \psi_\alpha^\mathrm{T}\cdot H_e\psi_{\alpha'}\big) \Big), \\
G_{\alpha\alpha'}^2 & = \tfrac{1}{2} \Big(R_{21} \big( \psi_\alpha^\mathrm{T}\cdot H_h \psi_{\alpha'}\big) + R_{22} \big( \psi_\alpha^\mathrm{T}\cdot H_e \psi_{\alpha'}\big)\Big).
\end{aligned}
\]

Note that, since $R_{21} = \tau R_{22}$ and $\widehat{H}_e \psi_2 = 0$, the only non-zero coefficient of the heat mode coupling matrix ~$G^2$  is $G^2_{11}$, which can be written more concisely as 
\[
G^2_{11} = -R_{22} \big(\psi_1^\mathrm{T} \cdot\widehat{H}_e \psi_1\big) = -  R_{22}\frac{(\partial_h \tau-\tau \partial_e \tau)^2}{Z_1^2} = -\left(\frac{c}{2}\right)^2 \frac{1}{\widetilde{Z}_2 Z_1^2} = - \frac1\beta \sqrt{-\frac{c}{2\Gamma}} <0.
\]
On the other hand, there seems to be no simplified expression for the sound mode coupling matrix ~$G^1$ and, \emph{a priori}, all entries $G^1_{\alpha \alpha'}$ are non-zero. A straighforward computation shows that 
\begin{equation}
  \label{eq:G^1_11}
  G^1_{11} = \frac{c}{2 Z_1^2 \widetilde{Z}_1} \left(\partial_h - \tau \partial_e\right)^2\tau,
\end{equation}
which has no definite sign, in general. 
\subsection{Specific potentials}
\label{sec:expressions_G_specific_pot}

There are simplifications for the expression of the components of the matrix $G^1$ for specific potentials such as the Kac-van Moerbeke potential~\eqref{eq:KVM_pot}. In the latter case, a simple computation based on the identity $V'(\eta) = -\kappa V(\eta) + \eta$ shows that 
\[
\tau = \kappa e - h,
\]
so that $c = -2(1+\kappa \tau)$, $H_h = 0$, 
\[
H_e = -2\begin{pmatrix} 1 & -\kappa \\ -\kappa & \kappa^2 \end{pmatrix}, 
\qquad
R = \frac{1}{\sqrt{1+\kappa\tau}}\begin{pmatrix} 
-\sqrt{\beta} & \kappa \sqrt{\beta} \\
\tau/\sqrt{\Gamma} & 1/\sqrt{\Gamma} \\ 
\end{pmatrix}.
\]
In addition, 
\[
\psi_1 = \sqrt{\frac{1}{\beta(1+\kappa\tau)}} \begin{pmatrix} 1 \\ -\tau \end{pmatrix},
\qquad 
\psi_2 = \sqrt{\frac{\Gamma}{1+\kappa\tau}} \begin{pmatrix} \kappa \\ 1 \end{pmatrix},
\]
so that $H_e \psi_2 = 0$. The only non-zero coefficient of $G^1$ therefore is $G_{11}^1$, which reads
\[
G^1_{11} = -\kappa\sqrt{\frac{1+\kappa\tau}{\beta}}.
\]
Note that the harmonic potential $V(\eta) = \frac{\eta^2}{2}$ is obtained from the KvM potential~\eqref{eq:KVM_pot} in the limit $\kappa\to 0$. Hence also the coupling matrices are obtained in the same limit, implying that 
$G^1 = 0$ for the harmonic potential.
\subsection{Coupling matrices for the numerically simulated systems}

Recall that we choose  $\tau = 1$ and $\beta = 2$ in both cases. For the FPU potential~\eqref{eq:FPU_pot} with $a=2$, we obtain $c = -5.28$,
\[
R = \begin{pmatrix}
-0.401 & 1.90 \\
2.55 & 2.55 \\ 
\end{pmatrix}, 
\qquad
R^{-1} = \begin{pmatrix}
-0.435 & 0.323 \\ 
0.435 & 0.0683 \\
\end{pmatrix}
\]
and 
\begin{equation}
\label{eq:G_for_FPU}
G^1 = \begin{pmatrix} 
  -2.23 & 0.431 \\ 
  0.431 & 0.333 \\
\end{pmatrix},
\qquad
G^2 = \begin{pmatrix} 
  -3.37 & 0 \\
  0 & 0\\
\end{pmatrix}.
\end{equation}
For the KvM potential~\eqref{eq:KVM_pot} with $\kappa=1$, we obtain  $c=-4$,
\[
R = \begin{pmatrix}
  - 1 & 1 \\  
  2.72 & 2.72 \\  
\end{pmatrix}, 
\qquad
R^{-1} = \begin{pmatrix}
  -1/2 & 0.184 \\
  1/2 & 0.184 \\
\end{pmatrix}
\]
and 
\begin{equation}
\label{eq:G_for_KVM}
G^1 = \begin{pmatrix} 
  -1 & 0 \\
  0 & 0 \\
\end{pmatrix},
\qquad
G^2 = \begin{pmatrix} 
  -2.719 & 0 \\
  0 & 0\\
\end{pmatrix}.
\end{equation}

%------------------ fin appendice -------------------
\end{appendix}

%---------------- bibliography ------------------------
%\bibliographystyle{plain}
%\bibliography{biblio}

\begin{thebibliography}{99}

\bibitem{ErHa76} M.H.~Ernst, E.H.~Hauge, and J.M.J~van~Leeuwen, 
Asymptotic time behavior of correlation functions. II. Kinetic and potential terms. J. Stat. Phys. \textbf{15}, 7--22 (1976).

\bibitem{FoNe77} D.~Forster, D.R.~Nelson, and M.J.~Stephen, Large-distance and long-time properties of a randomly stirred fluid. Phys. Rev. A \textbf{16}, 732--749 (1977).

\bibitem{Livi97} S.~Lepri, R.~Livi, and A.~Politi, Heat conduction in chains of nonlinear oscillators. Phys. Rev. Lett. \textbf{78}, 1896--1899 (1997). 

\bibitem{LeLi03} S.~Lepri, R.~Livi, and A.~Politi, Thermal conduction in classical low-dimensional lattices. Phys. Rep. \textbf{377}, 1--80 (2003).

\bibitem{Dh08} A.~Dhar, Heat transport in low-dimensional systems. Adv. Physics \textbf{57}, 457--537 (2008).

\bibitem{vB12} H.~van~Beijeren, Exact results for anomalous transport in one-dimensional Hamiltonian systems. Phys. Rev. Lett. \textbf{108}, 180601 (2012).

\bibitem{Spohn14} H.~Spohn, Nonlinear fluctuating hydrodynamics for anharmonic chains. J.~Stat. Phys. \textbf{154}, 1191--1227 (2014).

\bibitem{KPZ} M.~Kardar, G.~Parisi, and Y.-C.~Zhang, Dynamic scaling of growing interfaces. Phys. Rev. Lett.
\textbf{56},  889--892 (1986).

\bibitem{Co12} I.~Corwin, The Kardar-Parisi-Zhang equation and universality class. Random Matrices: Theory and Applications \textbf{1}, 113001 (2012). 

\bibitem{BG12} A.~Borodin and V.~Gorin, Lectures on integrable probability. \texttt{arXiv}: 1212.3351 (2012).

\bibitem{BP13} A.~Borodin and L.~Petrov, Integrable probability: from representation theory to Macdonald processes.
\texttt{arXiv}: 1310.8007 (2013).

\bibitem{QR13} J.~Quastel and D.~Remenik, Airy processes and variational problems. \texttt{arXiv}: 1301.0750 (2013).

\bibitem{BS12} C.~Bernardin and G.~Stoltz, Anomalous diffusion for a class of systems with two conserved quantities. Nonlinearity \textbf{25}, 1099--1133 (2012).

\bibitem{Ha13} M.~Hairer, Solving the KPZ equation. Annals Math. \textbf{178}, 559--664 (2013).

\bibitem{FuQu14} T. Funaki and J. Quastel, KPZ equation, its renormalization and invariant measures. \texttt{arXiv}:1407.7310 (2014).

\bibitem{BoCoFe14a} A.~Borodin, I.~Corwin, P.~Ferrari, and  B.~Vet\"{o}, Height fluctuations for the stationary KPZ equation. \texttt{arXiv}:1407.6977 (2014).

\bibitem{ImSa12} T.~Imamura and T.~Sasamoto, Stationary correlations for the 1D KPZ equation.
J.~Stat. Phys. \textbf{150}, 908--939 (2013).

\bibitem{Prhp} M. Pr\"ahofer, Exact scaling functions for one-dimensional stationary {KPZ} growth. \texttt{http://www-m5.ma.tum.de/KPZ}

\bibitem{PrSp04} M. Pr\"ahofer and H. Spohn, Exact scaling functions for one-dimensional stationary {KPZ} growth. J.~Stat. Phys. \textbf{115}, 255--279 (2004).

\bibitem{FeSp06} P.~Ferrari and H.~Spohn, Scaling limit for the space-time covariance of the stationary totally asymmetric simple exclusion process. Comm. Math. Phys. \textbf{265}, 1--44 (2006).

\bibitem{EK93}
D.~Erta\c{s} and M.~Kardar, Dynamic relaxation of drifting polymers: a
  phenomenological approach. Phys. Rev. E \textbf{48}, 1228--1245 (1993).

\bibitem{MS13} Ch.~B. Mendl and H.~Spohn, Dynamic correlators of Fermi-Pasta-Ulam chains and nonlinear fluctuating hydrodynamics. Phys. Rev. Lett. \textbf{111}, 230601 (2013).

\bibitem{BGJ14} C.~Bernardin, P.~Gon\c{c}alves, and M.~Jara, $3/4$-superdiffusion in a system of harmonic oscillators perturbed by a conservative noise. \texttt{arXiv}:1402.1562 (2014).

\bibitem{FeSS13} P.~Ferrari, T.~Sasamoto, and H.~Spohn,  Coupled Kardar-Parisi-Zhang equations in one dimension.
J. Stat. Phys. \textbf{153}, 377--399 (2013).

\bibitem{PSS14} V.~Popkov, J.~Schmidt, and G.~M. Sch\"utz, Superdiffusive modes in two-species driven diffusive systems. Phys. Rev. Lett. \textbf{112}, 200602 (2014).

\bibitem{PSS14a} V.~Popkov, J.~Schmidt, and G.~M. Sch\"utz, Universality classes in two-component driven diffusive systems. preprint (2014).

\bibitem{KuLa13} M.~Kulkarni and A.~Lamacraft, Finite-temperature dynamical structure factor of the one-dimensional Bose gas: From the Gross-Pitaevskii equation to the Kardar-Parisi-Zhang universality class of dynamical critical phenomena.
Phys.~Rev.~A \textbf{88}, 021603(R) (2013).

\bibitem{KuHuSp14} M.~Kulkarni, H.~Spohn and D.~Huse, Nonlinear fluctuating hydrodynamics for the 1D Bose gas, draft.

\bibitem{MeSp14b} Ch.~B. Mendl and H. Spohn, Nonlinear lattice Schr\"{o}dinger equation at low temperatures, in preparation.

\bibitem{KvM75} M.~Kac and P.~van Moerbeke, On an explicitly soluble system of nonlinear differential equations related to certain Toda lattices. Adv. Math. \textbf{16}, 160--169 (1975).

\bibitem{Toda88} M.~Toda, {\em Theory of Nonlinear Lattices (second enlarged edition)}, volume~20 of {\em Springer series in Solid-State Sciences} (Springer, 1988).

\bibitem{MS14} Ch.~B. Mendl and H.~Spohn, Equilibrium time-correlation functions for one-dimensional hard-point systems. Phys. Rev. E \textbf{90}, 012147 (2014).

\bibitem{DDSMS14} S.~G. Das, A.~Dhar, K.~Saito, Ch.~B. Mendl, and H.~Spohn, Numerical test of hydrodynamic fluctuation theory in the Fermi-Pasta-Ulam chain. Phys. Rev. E \textbf{90}, 012124 (2014).

\bibitem{St14} M. Straka, KPZ scaling in the one-dimensional FPU-model. Master Thesis, University of Florence, Italy (2013).

\bibitem{zwillinger}
D.~Zwillinger, {\em CRC Standard Mathematical Tables and Formulae (31st ed.)} (CRC Press, 2003).

\bibitem{JKO14} M.~Jara, T.~Komorowski, and S.~Olla, Superdiffusion of energy in a chain of harmonic oscillators with noise.
\texttt{arXiv}:1402.2988 (2014).

\bibitem{UZ99} V.~Uchaikin and V.~Zolotarev, {\em Chance and Stability. Stable Distributions and Applications}. Modern Probability and Statistics Series (De Gruyter, 1999).

\end{thebibliography}

%%%%%%%%%%%%%%%%%%%%%%%%%%%%%%%%%%%%%%%%%%%%%%%%%%%%%%%%%

\end{document}